\documentclass[12pt]{article}
\usepackage[dvips]{graphicx}
\usepackage{amsbsy}
\usepackage{amssymb}
\usepackage{float}
\usepackage{wrapfig}
\usepackage{color}

\newcommand{\beq}{\begin{equation}}
\newcommand{\eeq}{\end{equation}}
\newcommand{\bdm}{\begin{displaymath}}
\newcommand{\edm}{\end{displaymath}}
\newcommand{\beqr}{\begin{eqnarray}}
\newcommand{\eeqr}{\end{eqnarray}}
\newcommand{\beqrn}{\begin{eqnarray*}}
\newcommand{\eeqrn}{\end{eqnarray*}}

\begin{document}

\title{{\bf A Note on a Generalized AHM Model with Analytical Vortex Solutions}}

\author{A. Alonso Izquierdo$^\dag$, W. Garc\'{\i}a Fuertes$^*$,\\
J. Mateos Guilarte$^{\dag\dag}$}
\date{}
\maketitle
\begin{center}
$^{\dag}${{\it Departamento de Matem\'atica Aplicada, Facultad de Ciencias Agrarias y Ambientales,  Universidad de Salamanca,
E-37008 Salamanca, Spain.}}
\\ \vspace{0.2cm} $^*${{\it Departamento de F\'{\i}sica, Facultad de Ciencias,  Universidad de Oviedo,\\
E-33007 Oviedo, Spain.}}
\\ \vspace{0.2cm}
$^{\dag\dag}${{\it IUFFyM, Facultad de Ciencias,  Universidad de Salamanca,
E-37008 Salamanca, Spain.}}
\vskip1cm
\end{center}

\date{ }

\maketitle

\begin{abstract}
\noindent
We study topological vortex solutions in a generalized Abelian Higgs model with non-polynomial dielectric and potential functions. These quantities are chosen by requiring integrability of the self-dual limit of the theory for all values of the magnetic flux. All the vortex profiles are described by exact analytical expressions that solve the self-dual vortex equations. There is only a symmetry-breaking superconducting phase and the model sustains regular phenomenology. 
\end{abstract}
\bigskip


\medskip

\vfill\eject 
\section{Introduction}
Topological defects involving Higgs and/or gauge fields have aroused a widespread interest and are a central theme of research in several branches of physics, and also in mathematics \cite{Ra82,mansut}. The most prominent examples are the kinks of $\phi^4$ or sine-Gordon theories in 1+1 dimensions, the vortices of the Abelian Higgs model in 2+1 dimensions, the monopoles of the Georgi-Glashow model in 3+1 dimensions and the instantons of pure Yang-Mills theory in Euclidean four dimensional space. In all these cases, at least for some values of the parameters of the theory, there are bounds in energy or action leading the defects to obey first-order field equations, called Bogomolny or self-duality equations, in contrast with the usual second-order Euler-Lagrange ones. Vortices are, however, special in this respect in that there are not available exact solutions to these equations, while analytical expression for kinks, BPS monopoles and BSTP instantons are known. Although a well-defined procedure to obtain the coefficients of a series expansion of the fields has been developed \cite{deVeSch}, and a remarkable exact result for the leading term of the fields at large distance from the vortex core has been found\footnote{There is a small discrepancy of around 1.5\% between this theoretical prediction and the corresponding numerical coefficient obtained by de Vega and Schaposnik; a recent evaluation by high accuracy numerical methods \cite{ohashi} has concluded that the correct value is that given in \cite{deVeSch}.}\cite{tong}, no closed expressions of the scalar and vector fields of the Abelian Higgs model vortices have been brought to light. The situation changes, however, when vortices are considered on a curved manifold, where the metric can possibly depend on the scalar field, instead of on the plane: in these cases several examples of integrable vortex equations have been found and classified, see \cite{gudna} and references therein. From a different perspective, the Abelian Higgs model on a curved spacetime and with coupling to the gravitational field has been also thoroughly investigated, see \cite{VilShe} for a review or \cite{edery} for some recent new solutions obtained numerically.

On the other hand, in the spirit of treating Abelian Higgs systems as effective field theories in condensed matter physics or high energy physics, several variants of the original AHM have been developed. In particular, the inclusion of a dielectric function multiplying the Maxwell term in the Lagrangian \cite{coreanos}, or of a metric in scalar field space, making the theory a non linear sigma model \cite{lohe}, or a combination of both extensions \cite{ba12}, have been studied in several situations, see for instance \cite{nosotros1,nosotros2,nivi, nosotros3}. An aspect of these generalizations that has been recently investigated is the possibility of obtaining analytical solutions for vortices. Thus, in \cite{casanaref40} the non-linear sigma model with dielectric function and $\phi^4$ or $\phi^6$ potentials was considered, and by positing some exact expression for the scalar field of the vortex, it was verified that there is a complete analytical solution compatible with a well-behaved form of the dielectric function. Other forms of the scalar field leading to analytical solutions in this class of models were found in \cite{ramadham}, in some cases relaxing the usual requirement that the dielectric function is positive definite for all field values. The paper \cite{bazref25} uses a generalized model to address the issue of finding vorticial solutions of compacton type in 2+1 dimensions, and includes, along with numerical ones, some analytic solutions which arise when a parameter governing the dielectric function is very large. The same situation occurs in other models analyzed in \cite{bazref42}, this time in the context of a general formalism leading to vortex obeying first-order equations. Finally, a procedure of broad applicability for obtaining analytical vortices, based on the stipulation of a definite dependence between the scalar and gauge fields, was proposed in \cite{baz1801}, and applied successfully to find solutions in several models with dielectric function and scalar field metric. 

A common feature of previous works is that the analytical vortex solutions found correspond mostly to the case of $n=1$ vorticity, or minimal quantized magnetic flux. Although cases with $n>1$ were considered, for instance, in \cite{casanaref40} and \cite{ramadham}, the corresponding dielectric functions were computed as a function of the radial coordinate $r$ in the vortex plane, instead of a function of the scalar field, with the results depending on $n$. It seems thus that analytical vortices with different vorticities do belong to different theories of the same type, not to the same theory. The same situation happens with the procedure proposed in \cite{baz1801}: the relationship between scalar and gauge fields has to be adjusted in such way that the finite energy boundary conditions for the gauge field are satisfied, and this leads, in general, to a dependence of the dielectric function and potential on the vorticity. It is thus desirable to find a well definite generalized model in which vortices of all vorticities are analytical. In this note we exhibit one such model, which differs of previously known examples in two respects: i) Although the model contains analytical vortices of any vorticity, these arise into one unique theory, i.e., the dielectric function and potential are fixed and independent of $n$; ii) The analiticity of vortices is not limited to the case of cylindrical symmetry, configurations corresponding to separated vortices are given by exact formulas too. The price to be paid is that the theory is non-polynomial, but this is not an important drawback given that the model is to be considered as an effective theory. Non-polynomials potentials were also considered before, for instance, in \cite{bazref42} for Chern-Simons-Higgs vortices. We will present the model and its solutions in the next two sections and then devote a third one to study its character as effective theory and some phenomenology.
\section{The model and cylindrically symmetric vortices}
The generalized Abelian Higgs model with dielectric function \cite{coreanos} is given by the following Lagrangian density:
\beq
{\cal L}_{(\phi,V_\alpha)}=-\frac{1}{4 \mu\left(\frac{|\phi|}{v}\right)} V_{\alpha\beta} V^{\alpha\beta} +\nabla_\alpha\phi^* \nabla^\alpha\phi-\frac{\lambda}{2}\mu\left(\frac{|\phi|}{v}\right)\left(|\phi|^2-v^2\right)^2,\label{moddimen}
\eeq
where, as in \cite{and20}, we have chosen to express the theory in terms of the inverse dielectric function $\mu\left(\frac{|\phi|}{v}\right)$ rather than using the true dielectric one $H=\frac{1}{\mu}$. We work in 3+1 dimensions with spacetime coordinates $y^\alpha$, $\alpha=0,1,2,3$;  the metric signature is $(+,-,-,-)$ and the Maxwell tensor and covariant derivatives are $V_{\alpha\beta}=\frac{\partial V_\beta}{\partial y^\alpha}-\frac{\partial V_\alpha}{\partial y^\beta}$, $\nabla_\alpha\phi=\frac{\partial \phi}{\partial y^\alpha}- i q V_\alpha \phi$. Notice that in our conventions a positive value of $q$ corresponds to negative electric charge. The fields $\phi$ and $V_\alpha$ and the vacuum expectation value $v$ have mass dimensions one, while the couplings $q$ and $\lambda$ and the function $\mu$ are dimensionless. This is the physical model in natural units. However, for our purposes it is more convenient to avoid dimensionful quantities and to apply the rescaling
\bdm
q v y^\alpha=x^\alpha\hspace{2cm} V_\alpha=v A_\alpha\hspace{2cm} \phi=v\varphi
\edm
so that ${\cal L}_{(\phi,V_\alpha)}=q^2 v^4 {\cal L}$ with a new Lagrangian density
\beq
{\cal L}=-\frac{1}{4 \mu(|\varphi|)} F_{\alpha\beta} F^{\alpha\beta} +D_\alpha\varphi^* D^\alpha\varphi-\frac{\lambda}{2 q^2}\mu(|\varphi|)\left(|\varphi|^2-1\right)^2,\label{lagr}
\eeq
where $F_{\alpha\beta}=\partial_\alpha A_\beta-\partial_\beta A_\alpha$ and $D_\alpha\varphi=\partial_\alpha\varphi-i A_\alpha\varphi$ (with $\partial_\alpha=\frac{\partial}{\partial x^\alpha}$), and all fields and couplings are now dimensionless. We will be interested in static and $x^3$-independent configurations. Thus, the dimensionless tension, or energy per unit length along the third axis, is 
\beq
E= \int d^2x\left\{\frac{1}{2 \mu(|\varphi|)} F_{12}^2+D_k\varphi^* D_k\varphi+\frac{\lambda}{2 q^2}\mu(|\varphi|)\left(|\varphi|^2-1\right)^2\right\}\label{ener}
\eeq
where latin indices take the values 1 and 2. The static Euler-Lagrange equations extracted from (\ref{ener}) are
\beqr
\partial_k\left[\frac{1}{\mu(|\varphi|)} F^{kj} \right]&=&-i\left(\varphi^* D^j \varphi-\varphi D^j\varphi^*\right)\label{el1}\\
D_k D^k \varphi&=&-\frac{\partial}{\partial\varphi^*}\left\{\frac{1}{2 \mu(|\varphi|)} F_{12}^2+\frac{\lambda}{2 q^2}\label{el2}\mu(|\varphi|)\left(|\varphi|^2-1\right)^2\right\}.
\eeqr
We will assume that $\mu(|\varphi|)$ is semidefinite positive and non-vanishing for $|\varphi|=1$. Thus, on account of (\ref{ener}) the scalar field of configurations with finite tension should go to the vacuum orbit $|\varphi|=1$ for $|\vec{x}|\rightarrow\infty$. This implies that by means of a gauge transformation we can write $\varphi(\theta)=e^{i n \theta}$ at infinity, where $n$ is an integer and $\theta$ is the polar angle in the $(x^1-x^2)$-plane. But since the quadratic term in covariant derivatives has also to vanish asymptotically, we should require as well that $A_k=n\partial_k\theta$ for $|\vec{x}|\rightarrow\infty$. This means that the magnetic flux is quantized\footnote{We will refer loosely to $F_{12}$ and $\int d^2x F_{12}$ as the magnetic field and magnetic flux, although really the third component of the magnetic field is $B=-F_{12}$.}: $\Phi_M=\int d^2x F_{12}=\oint_{|\vec{x}|=\infty} A_k dx^k=2\pi n$. Therefore, the space of configurations of finite tension splits into topological sectors labeled by the topological index $n\in{\mathbb Z}$.

As in the standard Abelian Higgs model, the tension (\ref{ener}) is amenable to a splitting into squares plus a remnant \cite{coreanos}
\beqr
E&=&\int d^2x\left\{\frac{1}{2 \mu(|\varphi|)} \left(F_{12}\pm\frac{\sqrt{\lambda}}{q}\mu(|\varphi|)\left(|\varphi|^2-1\right)\right)^2+\left|D_1\varphi\pm i D_2\varphi\right|^2\right\}\nonumber\\ &\pm& \int d^2x F_{12}\left(|\varphi|^2\left(1-\frac{\sqrt{\lambda}}{q}\right)+1\right)\label{splitting}
\eeqr
such that, in the self-dual limit $\lambda=q^2$, the last term becomes a purely boundary contribution proportional to the magnetic flux. Therefore, the solutions of the first-order Bogomolny equations
\beqr
F_{12}&=&\pm\mu(|\varphi|)\left(1-|\varphi|^2\right)\label{bog1}\\
D_1\varphi&\pm& iD_2\varphi=0\label{bog2}
\eeqr
are minima of the tension in each topological sector and, thus, solutions also of the Euler-Lagrange equations. As we can read from (\ref{splitting}), the tension of these solutions is
\bdm
E=\pm \int d^2x F_{12}=\pm 2\pi n,
\edm
where $n>0$ for the upper sign and $n<0$ for the lower one. They are, respectively, the self-dual vortices and anti-vortices of the model.

Let us now specialize to the case of cylindrical symmetry. We work with radial $A_r$ and azimuthal $A_\theta$ gauge field components,  defined by $A_k=A_r \frac{\partial r}{\partial x^k}+A_\theta \frac{\partial \theta}{\partial x^k}$, $k=1,2$, and take an ansatz
\beq
A_r=0\hspace{1.5cm} A_\theta(r)=n-a(r)\hspace{1.5cm} \varphi(r,\theta)=g(r) e^{i n \theta},\label{ansatz}
\eeq
along with the boundary conditions needed to ensure finiteness of energy and regularity at the origin:
\beqr
g(0)=0&\hspace{3cm}&g(\infty)=1\label{bound1}\\
a(0)=n&\hspace{3cm}&a(\infty)=0\label{bound2}.
\eeqr
Thus, the first-order equations (\ref{bog1})-(\ref{bog2}) become
\beqr
\frac{1}{r}\frac{da}{dr}&=&\pm\mu(g)(g^2-1)\label{bograd1}\\
\frac{dg}{dr}&=&\pm\frac{ag}{r}\label{bograd2}
\eeqr
and the tension density ${\cal H}$, given by $E=\int d^2 x {\cal H}$, turns out to be
\beq
{\cal H}=\frac{1}{2\mu(g) r^2}\left(\frac{da}{dr}\right)^2+\left(\frac{dg}{dr}\right)^2+\left(\frac{ag}{r}\right)^2+\frac{1}{2}\mu(g)\left(g^2-1\right)^2.\label{edens}
\eeq
The Euler-Lagrange equations for cylindrically symmetric configurations can be written by substituting the ansatz (\ref{ansatz}) in (\ref{el1})-(\ref{el2}) or, alternatively, they can be derived from (\ref{edens}). They take the form
\beqr
\frac{d}{dr}\left(\frac{1}{r\mu(g)}\frac{da}{dr}\right)&=&\frac{2 a g^2}{r}\label{elrad1}\\
\frac{1}{r}\frac{d}{dr}\left(r\frac{dg}{dr}\right)&=&\frac{a^2g}{r^2}-\frac{1}{4 \mu^2(g) r^2}\left(\frac{da}{dr}\right)^2\frac{d\mu}{dg}+\frac{1}{4}\frac{d}{dg}\left(\mu(g)\left(g^2-1\right)^2\right).\label{elrad2}
\eeqr
It is not difficult to see that the Bogomolny equations (\ref{bograd1})-(\ref{bograd2}) do indeed imply (\ref{elrad1})-(\ref{elrad2}). Differentiation of (\ref{bograd1}) with respect to $r$, plus substitution of (\ref{bograd2}) in the right-hand side member, gives (\ref{elrad1}) directly. On the other hand, by means of (\ref{bograd2}) we can write the left-hand side member of (\ref{elrad2}) as $\frac{1}{r}\frac{d}{dr}\left(r\frac{dg}{dr}\right)=\pm \frac{1}{r}\frac{da}{dr} g+\frac{a^2g}{r^2}$, but then we see  by means of (\ref{bograd1}) that both the first term of this expression, and the two last terms of the right-hand side member of (\ref{elrad2}), are $\mu(g) g \left(g^2-1\right)$. Thus, they cancel and the second Euler-Lagrange equation follows. This means that the solutions of Bogomolny equations are true solutions of the theory even if the dielectric function is not positive definite, as we are assuming. However, in such a case they would not represent absolute minima of the tension in each topological sector and they could become unstable.

From now on we will focus on solutions with $n>0$, i.e., with positive magnetic flux. It is convenient \cite{jafftaub} to express the modulus of the scalar field as an exponential,  $g(r)=e^{u(r)}$. Equation (\ref{bograd2}) allows us to solve for the vector field in the form
\beq
a=r\frac{du}{dr},\label{afromu}
\eeq
and then, substituting in (\ref{bograd1}), all the problem reduce to the second-order ODE
\beq
r^2 \frac{d^2u}{dr^2}+r \frac{du}{dr}-r^2 \mu(u) (e^{2u}-1)=0\label{equ}
\eeq
which has to be solved with the boundary conditions coming from (\ref{bound1}):
\beq
u(0)=-\infty\hspace{3cm} u(\infty)=0. \label{boundu}
\eeq
Hence, if we are looking for an exact expression for the vortex fields, we have to choose a form of $\mu(u)$ such that it is possible to solve the system (\ref{equ})-(\ref{boundu}) analytically. In such a case, once $u(r)$ is found we obtain the gauge field from (\ref{afromu}) and we should check that (\ref{bound2}) is fulfilled. Apart of that, in order that the vortices are stable we should limit to semi-positive definite inverse dielectric functions, and additionally $\mu(|\varphi|)$ must be regular enough to not spoil spontaneous symmetry breaking when substituted in the potential term in (\ref{lagr}).

Let us then consider the following possibility:
\beq
\mu(|\varphi|)=\frac{\ln|\varphi|}{|\varphi|^2-1}.\label{choice}
\eeq
Both $\ln|\varphi| $ and $|\varphi|^2-1$ are negative if $|\varphi|<1$, and positive when $|\varphi|>1$. On the other hand, $\mu(|\varphi|)$ has  a regular limit for $|\varphi|=1$, namely  $\lim_{|\varphi|\rightarrow 1} \mu(|\varphi|)=\frac{1}{2}$. Therefore $\mu(|\varphi|)$ is positive definite and regular, as required. It decreases with $|\varphi|$ and vanishes for very high Higgs modulus, $\lim_{|\varphi|\rightarrow \infty}\mu(|\varphi|)=0$, blowing up to infinity for zero scalar field, $\lim_{|\varphi|\rightarrow 0}\mu(|\varphi|)=+\infty$. The potential in (\ref{lagr}) is
\beq
V(|\varphi|)=\frac{1}{2} \ln |\varphi|\left(|\varphi|^2-1\right)\label{poten}
\eeq
and the same balance of signs than before shows that it is positive for all values of $|\varphi|$, while $\lim_{|\varphi|\rightarrow 1} V(|\varphi|)=0$ and a vacuum orbit $|\varphi|=1$ exists guaranteeing spontaneous symmetry breaking. As the standard $|\varphi|^4$ potential of the usual Abelian Higgs model, $\lim_{|\varphi|\rightarrow \infty} V(|\varphi|)~=~+~\infty$, although in the present model the potential is not finite for zero field, $\lim_{|\varphi|\rightarrow 0} V(|\varphi|)=+\infty$.
The profiles of the inverse dielectric function (\ref{choice}) and the potential (\ref{poten}) are shown in Figure 1.
\begin{figure}[H]
\centering
\includegraphics[width=6cm]{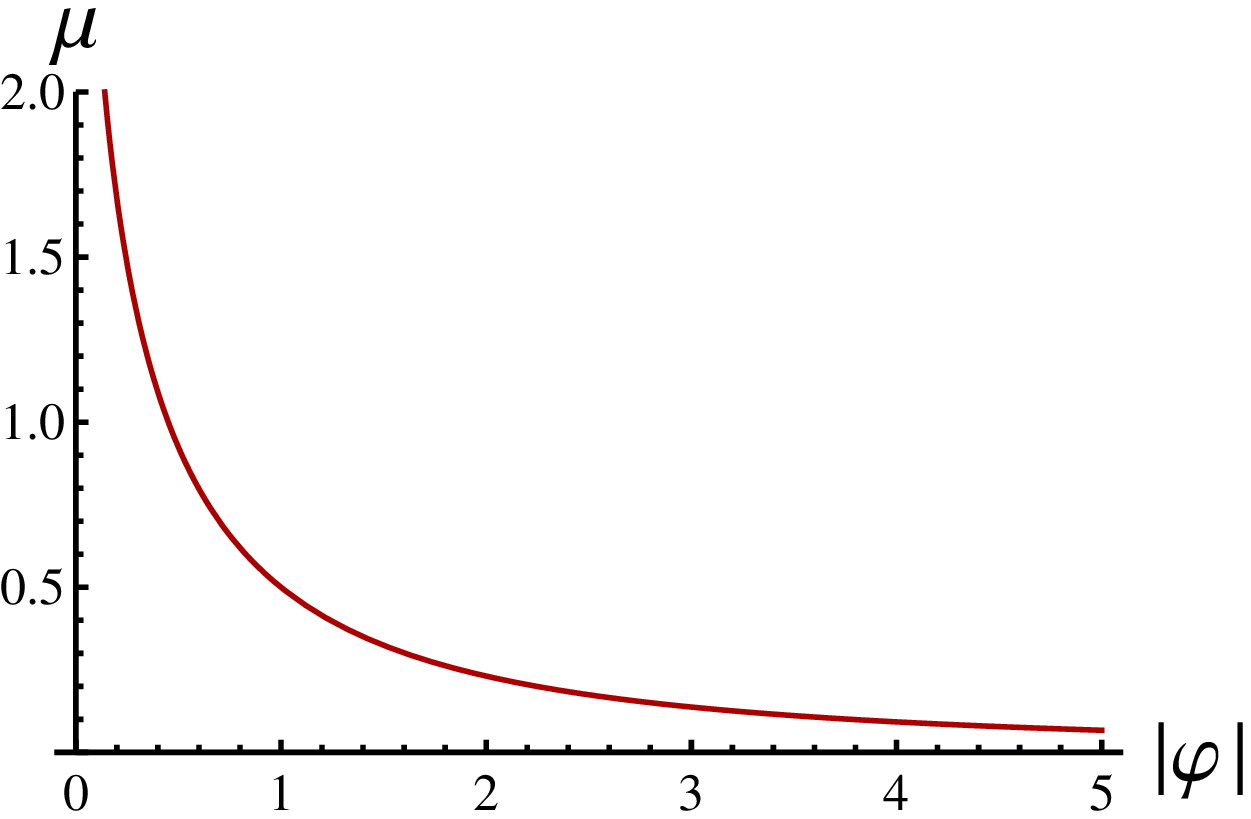}
\hspace{1cm}
\includegraphics[width=6cm]{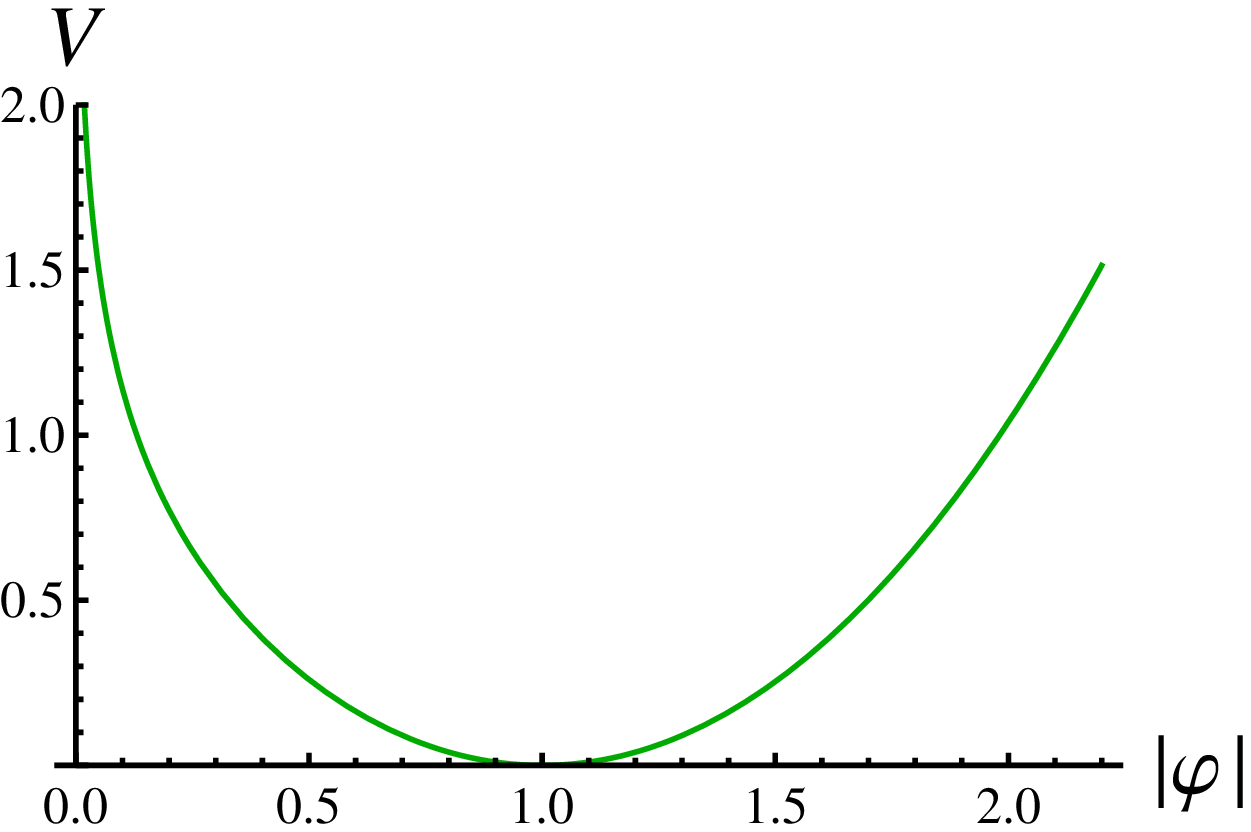}
\caption{The inverse dielectric function and potential for the model $(\ref{choice})-(\ref{poten})$. }
\label{fig1}
\end{figure}
\noindent With the choice (\ref{choice}), equation (\ref{equ}) reduces to the modified Bessel equation of zeroth order:
\beq
r^2 \frac{d^2u}{dr^2}+r \frac{du}{dr}-r^2 u=0,\label{eqbess}
\eeq
with general solution $u(r)=\alpha\, I_0(r)-\beta\, K_0(r)$. The boundary condition (\ref{boundu}) at infinity selects $\alpha=0$, while the behavior at the origin requires  $\beta>0$. In fact, due to (\ref{afromu}), we will need an integer $\beta$ in order to fulfill the boundary conditions for the vector field. Thus, choosing
\bdm
u(r)=-n K_0(r)
\edm
we obtain from (\ref{afromu})
\beq
a(r)=n r K_1(r)\label{ader}
\eeq
and, since $\lim_{z\rightarrow 0} z K_1(z)=1$ and $\lim_{z\rightarrow \infty} z K_1(z)=0$, the solution for $u(r)$ complies also with the boundary conditions (\ref{bound2}). The magnetic field can be obtained by means of standard Bessel function identities like $\frac{d K_n(z)}{dz}=-\frac{1}{2}\left(K_{n-1}(z)+K_{n+1}(z)\right)$ and $K_{n+1}(z)=K_{n-1}(z)+\frac{2n}{z} K_n(z)$, with the result
\beq
F_{12}=n K_0(r).\label{magnetic}
\eeq
The tension density, once the fields are substituted in (\ref{edens}) can also be computed to be of the form
\beq
{\cal H}(r)=n\left[K_0(r)-e^{-2 n K_0(r)}\left(K_0(r)-2 n K_1^2(r)\right)\right].\label{edensvort}
\eeq
It follows from (\ref{magnetic}) and (\ref{edensvort}) that both $F_{12}(r)$ and ${\cal H}(r)$ are divergent for $r\rightarrow 0$. Nevertheless, the important quantities for computing the magnetic flux and the tension of the cylindrically symmetric vortex are $r F_{12}(r)$ and $r {\cal H}(r)$, which are both regular at the origin because the divergence of $K_0(r)$ is logarithmic: $K_0(r)=-\ln r+(\ln 2-\gamma_E)+\frac{1}{4}(1-\gamma_E+\ln 2-\ln r) r^2+\ldots$, with $\gamma_E$ the Euler-Mascheroni constant, for small $r$. One can indeed check by doing the integrals that both the magnetic flux an the tension are finite and consistent with the Bogomolny bound, as it should be:
\bdm
\int_0^\infty dr r K_0(r)=1
\edm
and also
\bdm
\int_0^\infty dr r \left[K_0(r)-e^{-2 n K_0(r)}\left(K_0(r)-2 n K_1^2(r)\right)\right]=1,
\edm
because
\bdm
\int_0^\infty dr r K_0(r) e^{-2 n K_0(r)}=-r K_1(r)e^{-2 n K_0(r)}{\Big\vert}_0^\infty+2 n \int_0^\infty K_1^2(r)e^{-2 n K_0(r)}
\edm
and $\lim_{r\rightarrow 0} r K_1(r)e^{-2 n K_0(r)}=\lim_{r\rightarrow \infty} r K_1(r)e^{-2 n K_0(r)}=0$.

We present in Figure 2 the profiles of the scalar and gauge fields of the cylindrically symmetric vortices for several values of the topological index $n$, along with the corresponding densities of magnetic flux and tension, including the factor $r$. As one can see from the figure, the region around the vortex center with $\varphi\simeq 0$ gets wider as the vorticity increases, while at the same time the area with $A_\theta\simeq 0$ becomes stretched. Both $r F_{12}$ and $r {\cal H}$ are zero at the vortex center and have a maximum in the form of an annulus around it, with the top of the annulus flatter and wider for higher $n$ values. Because the boundary conditions are the same, the fields $g(r)$ and $a(r)$ of our solution and those of the the standard AHM vortices show a similar appearance, but there are some differences in the way the origin and infinity are approached. For instance, for the case $n=1$, the fields of our model near the origin are 
\beqrn
g(r)&=&\frac{e^{\gamma_E}}{2}  r+\frac{e^{\gamma_E}}{8} \left[\gamma_E -1 +\ln\left(\frac{r}{2}\right)\right] r^3+\ldots\\
a(r)&=&1+\frac{1}{4}\left[ 2\gamma_E-1+\ln\left(\frac{r}{2}\right)\right] r^2+\ldots,
\eeqrn
while in the self-dual limit of the AHM obtained by taking $\mu\left(\frac{|\phi|}{v}\right)=1$ in (\ref{moddimen}), one would find \cite{VilShe}
\beqrn
g(r)&=& \zeta \; r-\frac{\zeta }{4}r^3+\ldots\\
a(r)&=&1-\frac{1}{2} r^2+\frac{\zeta ^2}{4} r^4+\ldots,
\eeqrn
with $\zeta =0.8532$ and no logarithmic terms. On the other hand, the behavior for great $r$ is in both cases of the form
\beqrn
g(r)&\simeq&1-K_0(\varrho\,  r)\simeq 1-O(\sqrt{\frac{\pi}{2 \varrho\, r}}e^{-\varrho\, r})\\
a(r)&\simeq& r K_1(\varrho\, r)\simeq O(\sqrt{\frac{\pi r}{2\varrho}}e^{-\varrho\, r}),
\eeqrn
with $\varrho=1$ in our model, but $\varrho=\sqrt{2}$ in the AHM, signaling the fact that in this latter case the elementary bosons turn out to be $\sqrt{2}$ heavier.

We have obtained vortex solutions by solving equation (\ref{equ}) for $r>0$ and using boundary conditions (\ref{bound1})-(\ref{bound2}) at $r=0$. Alternatively, it is possible to extend (\ref{equ}) to the whole plane \cite{jafftaub}:  writing $\varphi(r,\theta)=e^{u(r)} e^{i n\theta}$ and substituting in (\ref{bog2}), we can solve for the vector field out of the origin in the form
\bdm
A_1=\partial_2 u+n\partial_1\theta,\hspace{2cm} A_2=-\partial_1 u+n \partial_2\theta
\edm
and thus the magnetic field picks a singular contribution
\beq
F_{12}=-(\partial_1^2+ \partial_2^2) u+n\varepsilon_{ij}\partial_i\partial_j\theta=-(\partial_1^2+ \partial_2^2) u+2\pi n\delta^{(2)}(\vec{r})\label{f12condelta}
\eeq
when the origin is included. Therefore, for the dielectric function (\ref{choice}), Bogomolny equation (\ref{bog1}) becomes
\beq
(\partial_1^2+ \partial_2^2-1)u=2\pi n\delta^{(2)}(\vec{r}), \label{ecua}
\eeq
and we recover our previous solution $u(r)=-n K_0(r)$, now because $-K_0(r)$ is the Green function of the Helmholtz operator (with -1 instead of +1): \beq
(\partial_1^2+ \partial_2^2-1)K_0(r)=-2\pi \delta^{(2)}(\vec{r}).\label{green}
\eeq
If instead we take $u(r)=-\beta K_0(r)$ with $\beta\neq n$, we obtain a solution in all the plane except the origin and, at the same time, the gauge field develops a singularity at this point, since in this case $A_\theta(\vec{0})=n-\beta$ is non-vanishing. The non-regularity of the gauge field manifests itself in a singulatity in the magnetic flux: from (\ref{f12condelta}) and (\ref{green})
\beq
F_{12}=\beta K_0(r)+2\pi (n-\beta)\delta^{(2)}(\vec{r})\label{f12mala}
\eeq
and a Dirac string emerges. Nevertheless, if we compute the total magnetic flux the result remains the same: like in regular solutions, the first term in (\ref{f12mala}) gives a contribution $2\pi\beta$, and adding the flux of the Dirac string we recover $\int d^2x F_{12}=2\pi n$. The energy is
\bdm
E=2\pi\beta+\frac{1}{2} \int_{r\leq \epsilon} d^2 x \frac{F_{12}^2}{\mu(|\varphi|)},
\edm
where the first term arises by substituting $n$ for $\beta$ in (\ref{edensvort}) and in the second we integrate the Maxwell term over a circle of infinitesimal radius $\epsilon$ around the origin to take care of the singularity in the magnetic flux. We can approximate $u(r)\simeq\beta\ln r$, and thus $\mu(|\varphi(r)|)\simeq-\beta\ln r$ in this circle, so that
\beqrn
\frac{1}{2}\int_{r\leq \epsilon} d^2 x \frac{F_{12}^2}{\mu(|\varphi|)}&=&-\frac{1}{2}\int_{r\leq \epsilon} d^2 x\frac{\beta^2 \ln^2 r-4\pi(n-\beta) \beta\ln r\; \delta^{(2)}(\vec{r})+4\pi^2 (n-\beta)^2 (\delta^{(2)}(\vec{r}))^2}{\beta\ln r}\\&=&-\frac{1}{2}\int_{r\leq \epsilon} d^2 x\left(\beta \ln r-4\pi(n-\beta) \delta^{(2)}(\vec{r})+4\pi^2 \frac{(n-\beta)^2}{\beta} \frac{\delta^{(2)}(\vec{r})}{\ln r} \delta^{(2)}(\vec{r})\right).
\eeqrn
Both the first and third terms integrate to zero, in the second case because $\frac{\delta^{(2)}(\vec{r})}{\ln r}$ is of the form $f(r) \delta^{(2)}(\vec{r})$ with $f(0)=0$. Thus
\bdm
E=2\pi \beta+2\pi(n-\beta)=2\pi n
\edm
and the singular pseudo solutions have in this model the same energy and flux that the true regular ones. This is in marked contrast with the standard Abelian Higgs Model, where a Dirac string singularity would carry infinite energy because the third term of the previous integral would lack the factor $\frac{1}{\ln r}$ coming from the dielectric function. Thus, in the present situation the configurations with $A_\theta(\vec{0})=n-\beta\neq 0$ are to be rejected uniquely on the basis that for them the gauge field is not well defined at the origin,
not because they have infinite energy.
\begin{figure}[t]
\centering
\includegraphics[width=6cm]{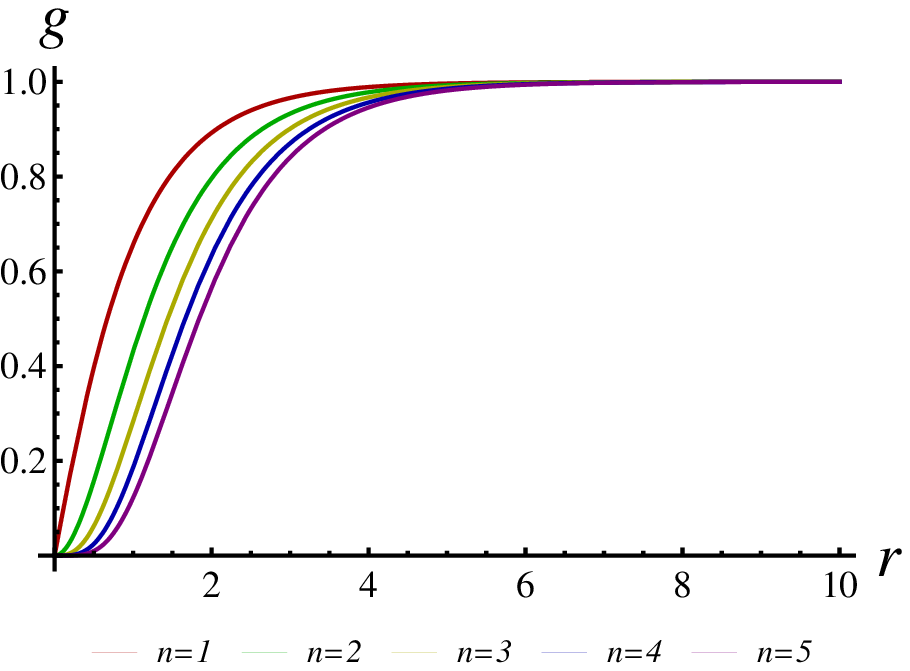}
\hspace{1cm}
\includegraphics[width=6cm]{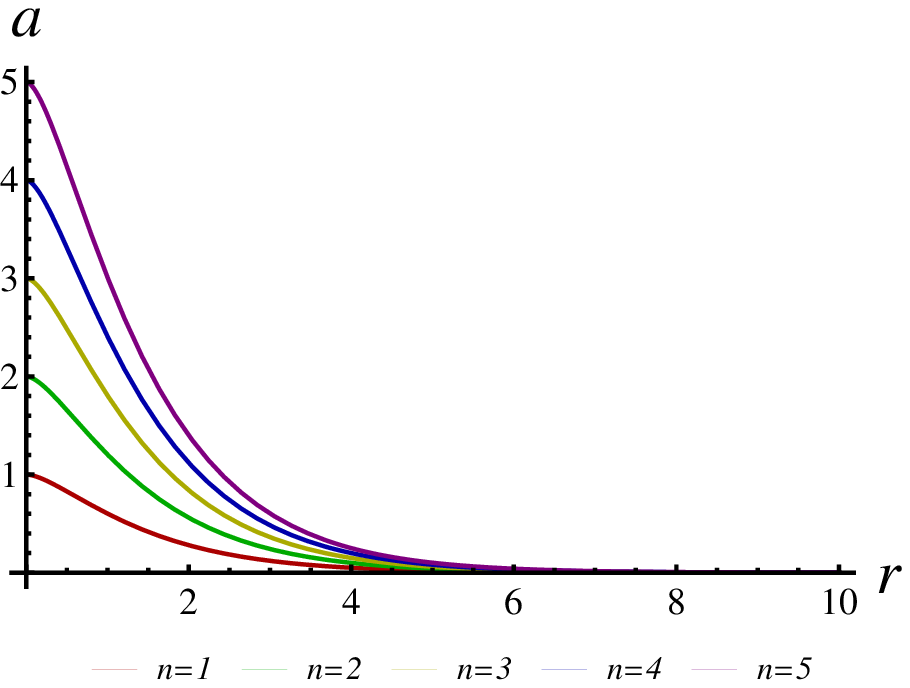}
\\
\includegraphics[width=6cm]{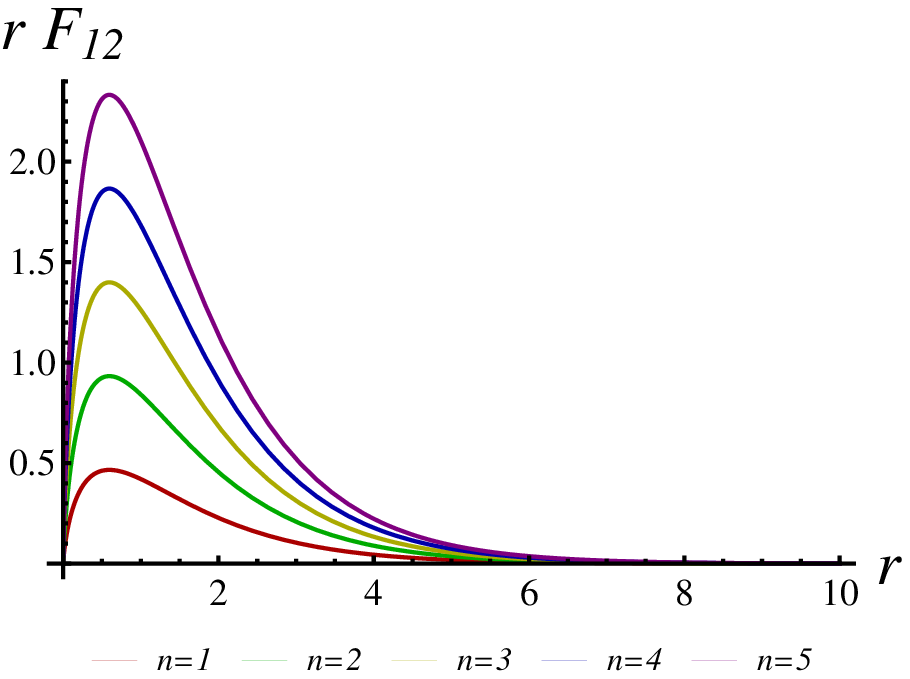}
\hspace{1cm}
\includegraphics[width=6cm]{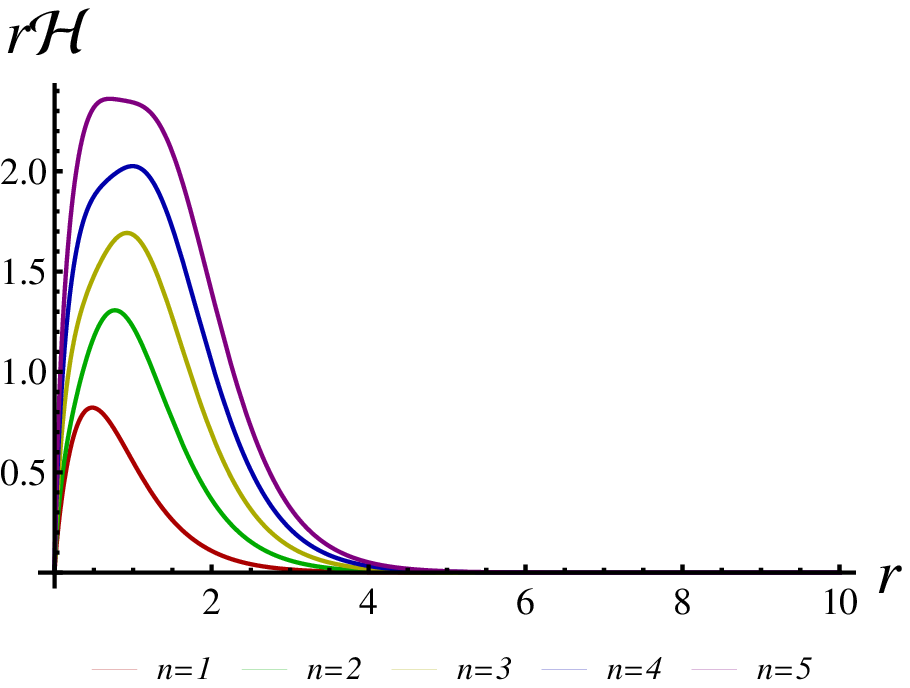}
\caption{Profiles of $g(r)$ and $a(r)$ and magnetic flux and tension densities for several vorticities. The red and magenta curves correspond, respectively, to $n=1$ and $n=5$.}
\label{fig2}
\end{figure}

Finally, let us comment that the election (\ref{choice}) can be slightly generalized by means of a new dimensionless positive coupling $\chi$ as
\bdm
\mu(|\varphi|)=\chi \frac{\ln|\varphi|}{|\varphi|^2-1},\hspace{1cm} \chi>0.
\edm
The profiles of $\mu(|\varphi|)$ and $V(|\varphi|)$ are qualitatively the same than in Figure 1, although now $\mu(1)=\frac{\chi}{2}$. The vortex fields are in this case of the form
\bdm
u(r)=-n K_0(\sqrt{\chi} r)\hspace{2cm}a(r)=n \sqrt{\chi} K_1(\sqrt{\chi} r)\hspace{2cm}F_{12}(r)=n\chi K_0(\sqrt{\chi} r),
\edm
and the tension density is (\ref{edensvort}) multiplied by a global factor $\chi$ and with the change $r\rightarrow \sqrt{\chi} r$. Thus, the effect of a small value of the coupling $\chi$ is to make the vortex core wider, with the magnetic field and energy density less concentrated around the center, whilst these magnitudes would be enhanced and confined into a narrow tube if $\chi$ were large. This can be seen in Figure 3, where the fields of the vortex with $n=2$ and several values of $\chi$ are displayed. In what follow, however, we will continue taking $\chi=1$.
\begin{figure}[t]
\centering
\includegraphics[width=6cm]{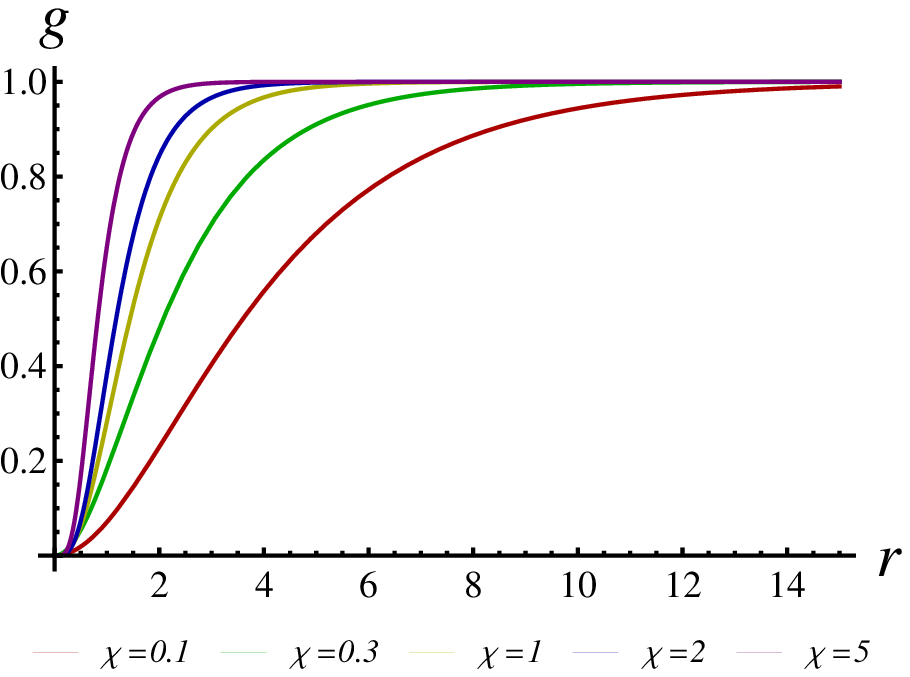}
\hspace{1cm}
\includegraphics[width=6cm]{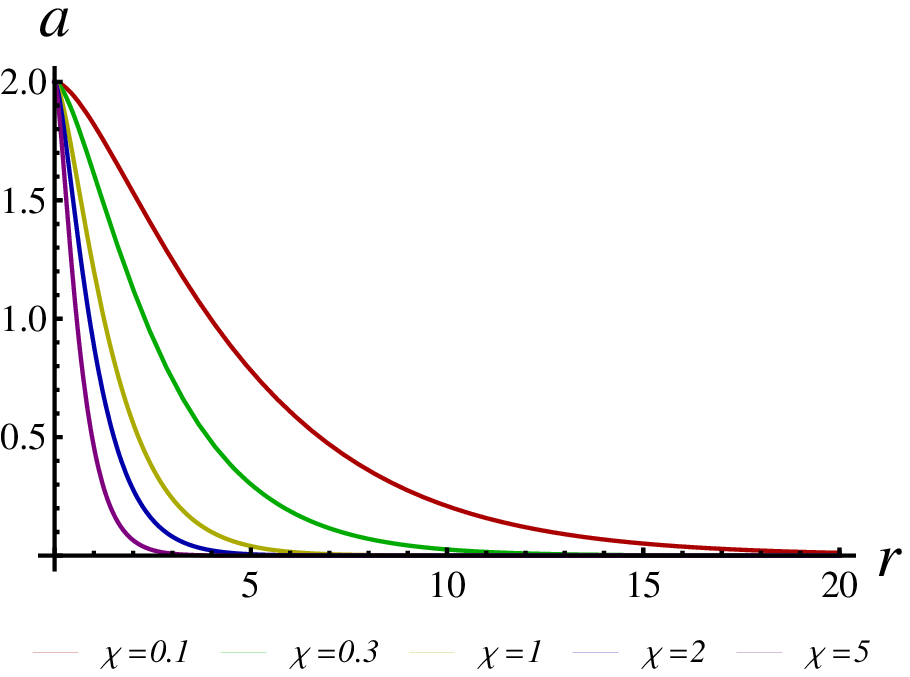}
\\
\includegraphics[width=6cm]{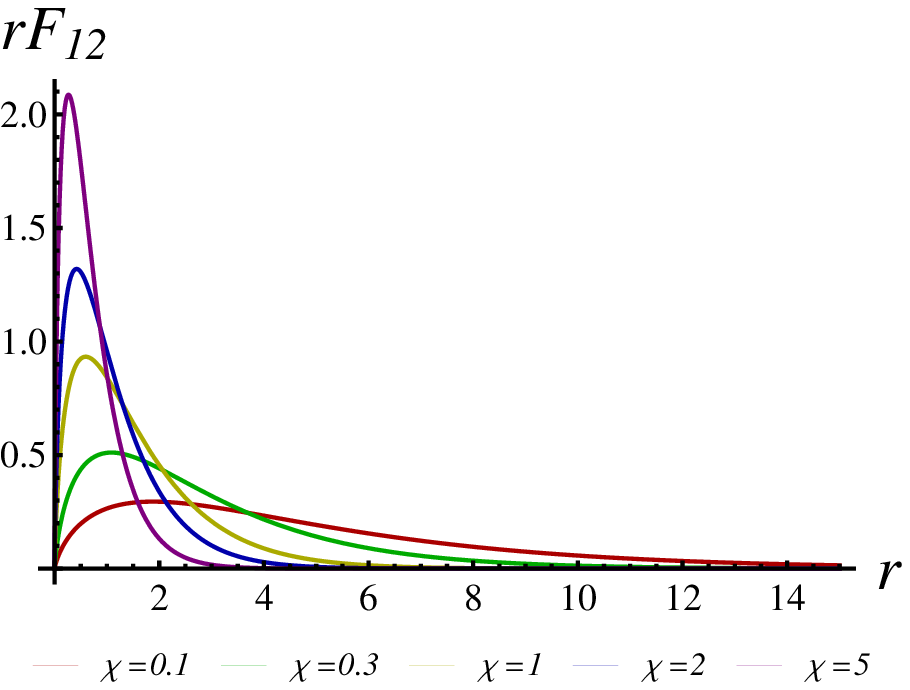}
\hspace{1cm}
\includegraphics[width=6cm]{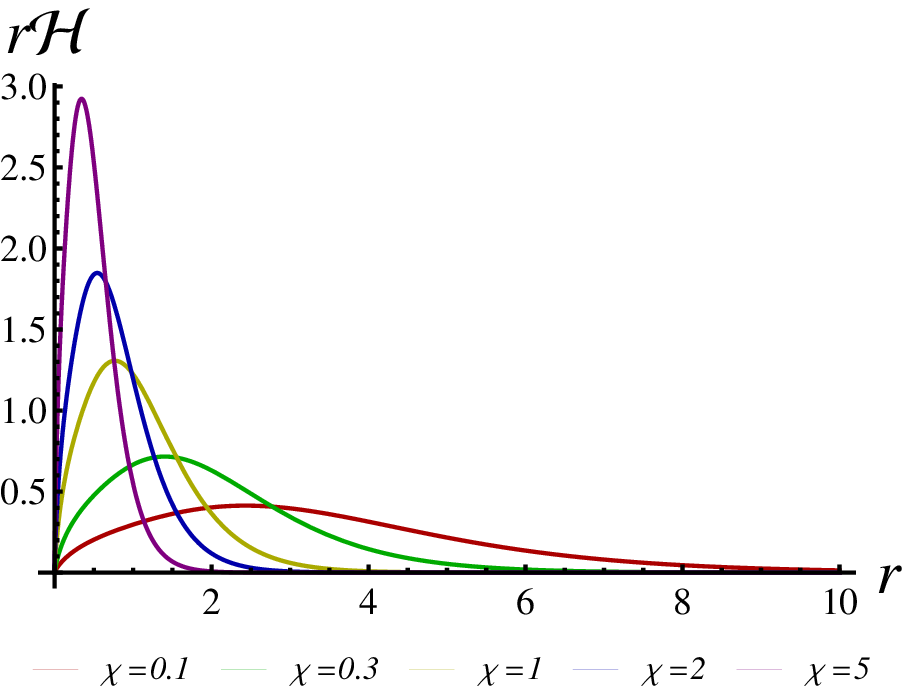}
\caption{Profiles of $g(r)$ and $a(r)$ and magnetic flux and tension densities for vorticity $n=2$ and $\chi=0.1,\  0.3,\  1,\  2\  {\rm and}\  5$. The red and magenta curves correspond, respectively, to $\chi=0.1$ and $\chi=5$.}
\label{fig3}
\end{figure}
\section{More general solutions}
As it is well known, cylindrically symmetric vortices are not the only topological solutions of the Abelian Higgs model with self-dual coupling. In fact, as proved in \cite{weinb} by means of index theorem techniques applied to the differential operator ruling the self-dual deformations of cylindrically symmetric vortices, and in \cite{jafftaub} through the construction of a functional whose critical points are in one-to-one correspondence with the solutions of the Bogomolny equations, the moduli space of solutions in the sector of topological index $n$ has dimension $2n$, and the general solution describes an equilibrium distribution of $n$ separated vortices centered in $n$ different points of the plane. The same situation occurs also in Abelian Higgs models with a dielectric functions, see for instance \cite{nosotros99} for the index computation in a generalized model related to the Chern-Simons-Higgs system. While the existence of assemblies of vortices distributed on the plane has generally to established by indirect means, we will show in this section that the model we are dealing with has the virtue of allowing an explicit analytical construction of the solutions reflecting the structure of moduli space. 

Thus, we are now looking for non-cylindrically symmetric solutions of the Bogomolny equations (\ref{bog1})-(\ref{bog2}) with the upper signs, and with the inverse dielectric function given by (\ref{choice}). We assume that the topological index is $n$ and that the scalar field has $n$ zeroes which are located at some given points $\vec{r}=\vec{r}_j$, $j=1,2,3,\ldots,n,$ of the plane. As in the cylindrically symmetric case, these zeroes correspond to the centers of the vortices. We proceed as in \cite{jafftaub} by taking the ansatz
\bdm
\varphi(\vec{r})=\exp\left[{u(\vec{r})+i \Omega(\vec{r})}\right]
\edm
where the gauge is chosen in the form
\beq
\Omega(\vec{r})=\sum_{k=1}^n\theta(\vec{r}-\vec{r}_k)\label{gaugeelect}
\eeq
with $\theta(\vec{s})$ the polar angle corresponding to position vector $\vec{s}$, i.e. $\theta(\vec{s})=\arctan(\frac{x^2}{x^1})$ for $\vec{s}~=~x^1 \vec{e}_1~+x^2 \vec{e}_2$. The boundary conditions for $u(\vec{r})$ are now
\beq
u(\vec{r}_k)=-\infty,\ \ k=1,2,\ldots,n;\hspace{2cm} u(\vec{r})|_{|\vec{r}|\rightarrow\infty}=0.\label{boundgen}
\eeq
Out of the vortex centers, equation (\ref{bog2}) 
gives the vector field components in terms of the modulus and phase of the scalar field:
\beq
A_1=\partial_2 u+\partial_1\Omega,\hspace{2cm} A_2=-\partial_1 u+\partial_2\Omega.\label{vector}
\eeq
Therefore, using (\ref{vector}) in (\ref{bog1}), we rewrite the equation for $u(\vec{r})$  in these points as
\beq
(\partial_1^2+\partial_2^2-1)u=0.\label{equgen}
\eeq
We have to solve (\ref{equgen}) with boundary conditions (\ref{boundgen}). We proceed directly in cartesian coordinates: using that $
\frac{dK_0(z)}{dz}=-K_1(z)$ and $\frac{dK_1(z)}{dz}=-K_0(z)-\frac{1}{z} K_1(z)$, we have
\bdm
\partial_j K_0(|\vec{r}|)=-\frac{x^j}{|\vec{r}|}K_1(|\vec{r}|),\hspace{1.5cm}
\partial_j^2 K_0(|\vec{r}|)=\frac{(x^j)^2}{|\vec{r}|^2}K_0(|\vec{r}|)+\frac{(x^j)^2-(\varepsilon_{jk}x^k)^2}{|\vec{r}|^3} K_1(|\vec{r}|),
\edm
and thus $(\partial_1^2+\partial_2^2-1)K_0(|\vec{r}|)=0$ for $\vec{r}\neq 0$. Hence, by choosing
\beq
u(\vec{r})=-\sum_{k=1}^n K_0(|\vec{r}-\vec{r}_k|)\label{solu}
\eeq
equation (\ref{equgen}) is satisfied out of the vortex centers. Also, since $K_0(0)=\infty$ and $K_0(\infty)=0$, (\ref{solu}) is consistent with the boundary conditions (\ref{boundgen}). Thus, the solution for the scalar field of the multivortex configuration is simply
\bdm
\varphi(\vec{r})=\prod_{k=1}^n e^{-K_0(|\vec{r}-\vec{r}_k|)} e^{i\theta(\vec{r}-\vec{r}_k)}.
\edm
With $u(\vec{r})$ found explicitly and the form of $\Omega(\vec{r})$ fixed by the gauge election, the vector field can be computed by means of (\ref{vector}), with the result
\beq
A_i(\vec{r})=-\varepsilon_{ij}\sum_{k=1}^n\left[\frac{x^j-x_k^j}{|\vec{r}-\vec{r}_k|^2}-\frac{x^j-x_k^j}{|\vec{r}-\vec{r}_k|}K_1(|\vec{r}-\vec{r}_k|)\right],\label{vec}
\eeq
where $\varepsilon_{ij}$ the antisymmetric symbol, $\varepsilon_{12}=1$. In fact, it is easy to see that this gauge field is a sum of fields of unit vorticity and, therefore, it is regular at the centers of the vortices. The term of (\ref{vec}) corresponding, for instance, to the vortex located at $\vec{r}=\vec{r}_1$ is 
\beqrn
A_1^{(1)}(\vec{r})&=&-\frac{x^2-x_1^2}{|\vec{r}-\vec{r}_1|^2}+\frac{x^2-x_1^2}{|\vec{r}-\vec{r}_1|}K_1(|\vec{r}-\vec{r}_1|)=-\frac{\sin\theta_1}{r_1} (1-r_1 K_1(r_1))\\
A_2^{(1)}(\vec{r})&=&\frac{x^1-x_1^1}{|\vec{r}-\vec{r}_1|^2}-\frac{x^1-x_1^1}{|\vec{r}-\vec{r}_1|}K_1(|\vec{r}-\vec{r}_1|)=\frac{\cos\theta_1}{r_1} (1-r_1 K_1(r_1))
\eeqrn
where we denote $\theta_1=\theta(\vec{r}-\vec{r}_1)$ and $r_1=|\vec{r}-\vec{r}_1|$. Going to the polar components of the gauge fields, as they were defined in Section 2,  we conclude that this corresponds to $A_{r_1}^{(1)}=0$ and $A_{\theta_1}^{(1)}=1- r_1 K_1(r_1)$, which coincides with (\ref{ansatz})-(\ref{ader}) for $n=1$. Hence, $A_{\theta_1}^{(1)}$ goes to zero at the center of the vortex, as it should do.

Since (\ref{vec}) is a sum of unit vorticity vector fields, so is the magnetic field. We can see this in Cartesian coordinates: out of the vortex centers and for $i\neq j$, we obtain
\bdm
\partial_i A_j=\sum_{k=1}^n\left[\varepsilon_{ij}\left(\frac{1}{|\vec{r}-\vec{r}_k|^2}-2\frac{(x^i-x^i_k)^2}{|\vec{r}-\vec{r}_k|^4}-\frac{1}{|\vec{r}-\vec{r}_k|}K_1(|\vec{r}-\vec{r}_k|)+\frac{(x^i-x^i_k)^2}{|\vec{r}-\vec{r}_k|^2}K_2(|\vec{r}-\vec{r}_k|)\right)\right],
\edm
where there is no sum in $i$ and $j$, and we have used $\frac{d}{dz}\left(\frac{K_1(z)}{z}\right)=-\frac{K_2(z)}{z}$. Thus
\bdm
F_{12}=\sum_{k=1}^n\left[-\frac{2}{|\vec{r}-\vec{r}_k|}K_1(|\vec{r}-\vec{r}_k|)+K_2(|\vec{r}-\vec{r}_k|)\right]
\edm
but $-\frac{2}{z} K_1(z)+K_2(z)=K_0(z)$, and then
\bdm
F_{12}=\sum_{k=1}^n K_0(|\vec{r}-\vec{r}_k|),
\edm
which is consistent with (\ref{bog1}). As in the cylindrically symmetric case, the magnetic field diverges at the vortex core, but the magnetic flux is nevertheless finite. 

As an illustration, we present in Figure 4 the splitting of a cylindrically symmetric vortex with $n=2$ into two separated vortices, with the distance between centers increasing from 0 to 5 units along the $x^1$ axis. The figure shows the modulus of the scalar field of the vortices and, for convenience, the vertical axis has been inverted, i.e. the summit of the hills are the vortex centers with $\varphi=0$ and the surrounding flat landscape corresponds to $|\varphi|=1$.
\begin{figure}[t]
\centering
\includegraphics[width=5cm]{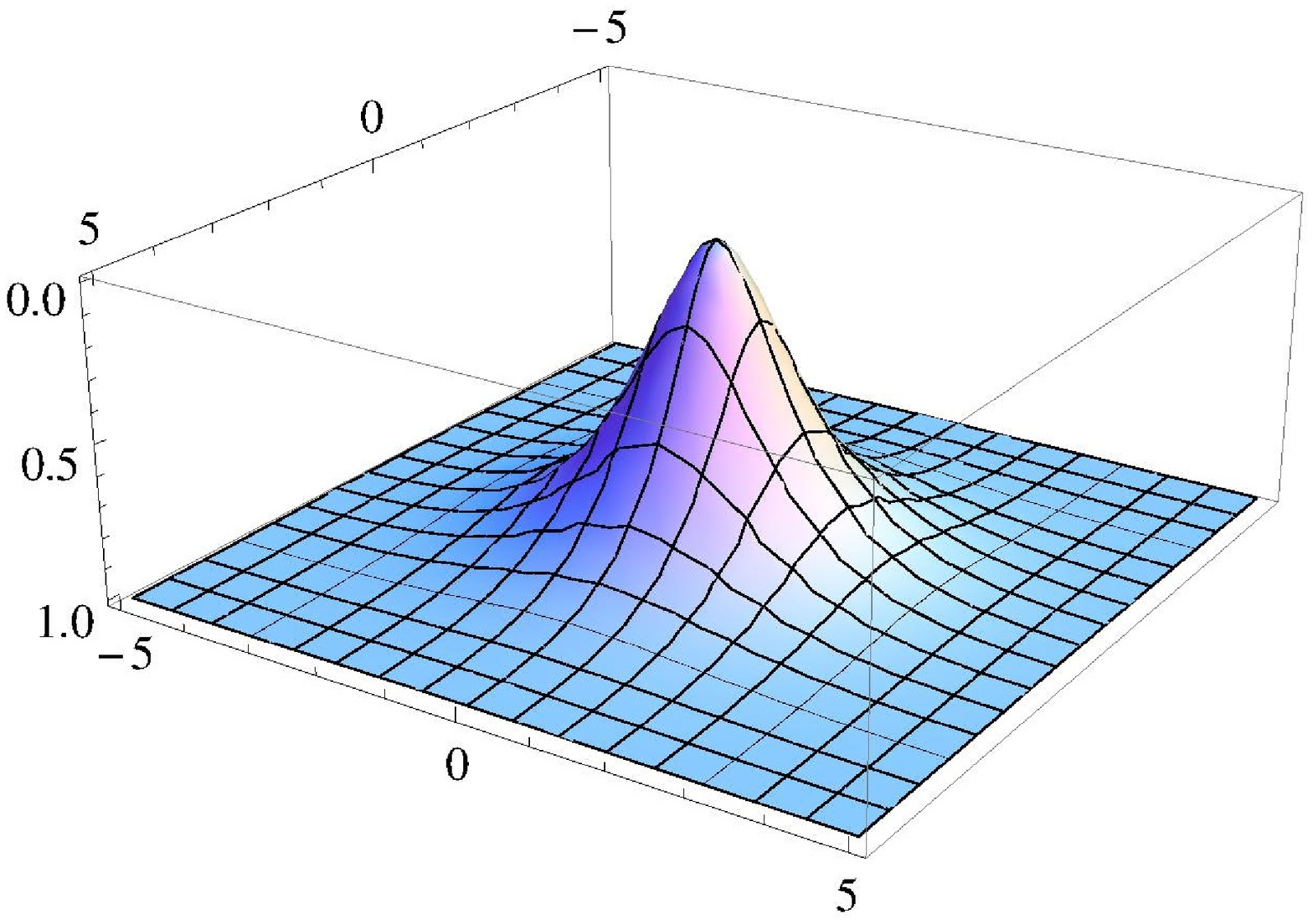}
\hspace{0.2cm}
\includegraphics[width=5cm]{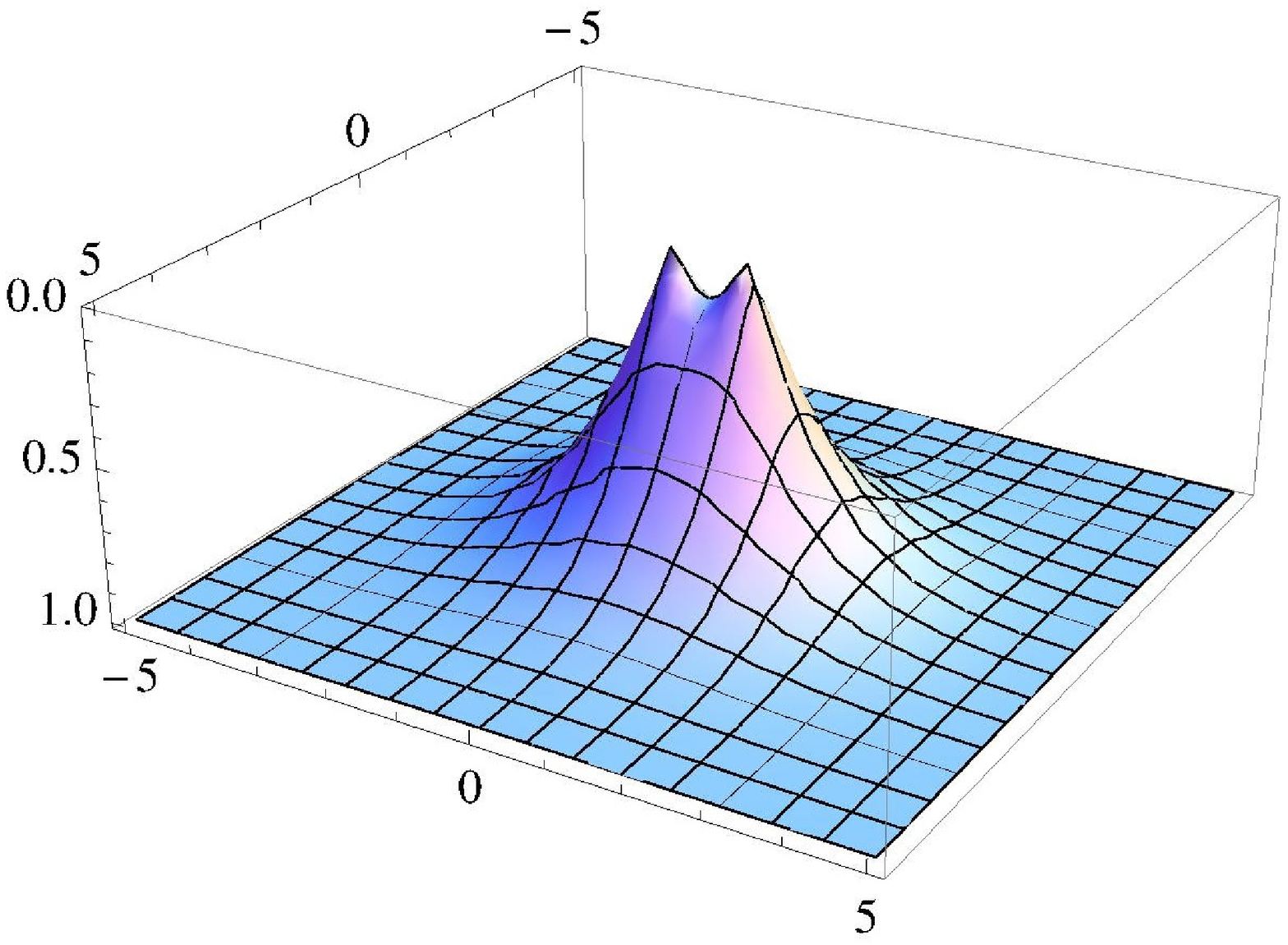}
\hspace{0.2cm}
\includegraphics[width=5cm]{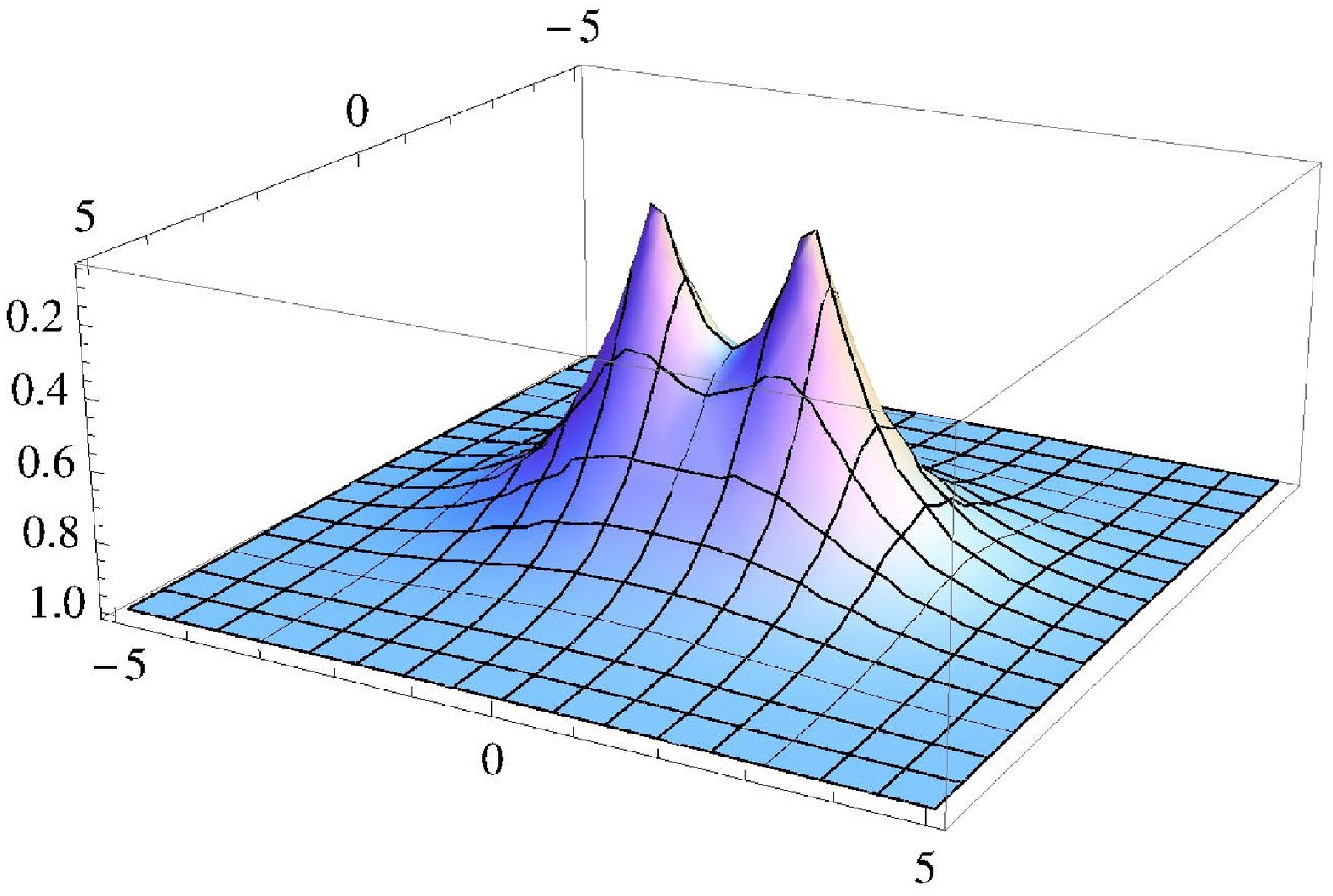}\\
\hspace{0.2cm}
\includegraphics[width=5cm]{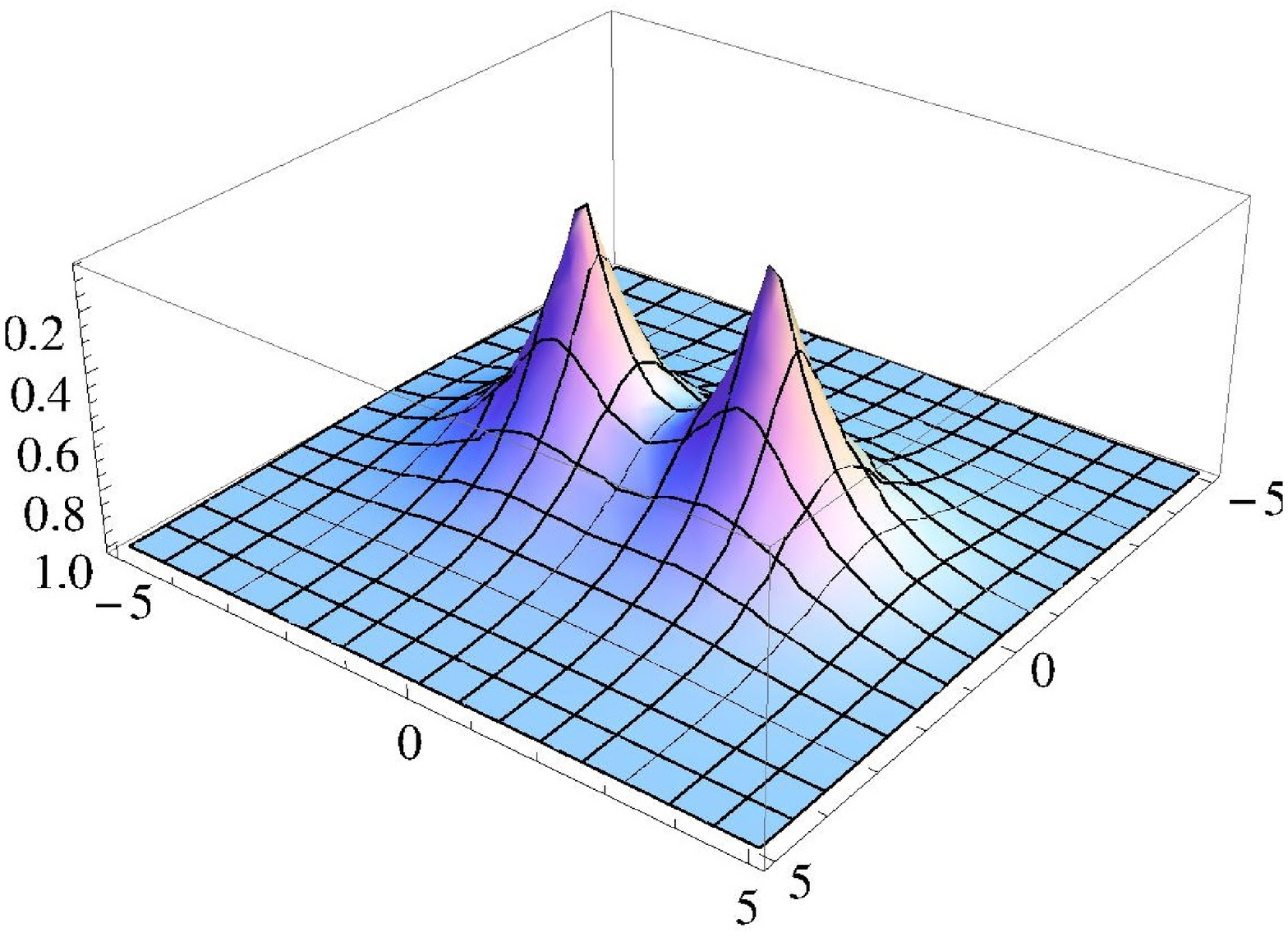}
\hspace{0.2cm}
\includegraphics[width=5cm]{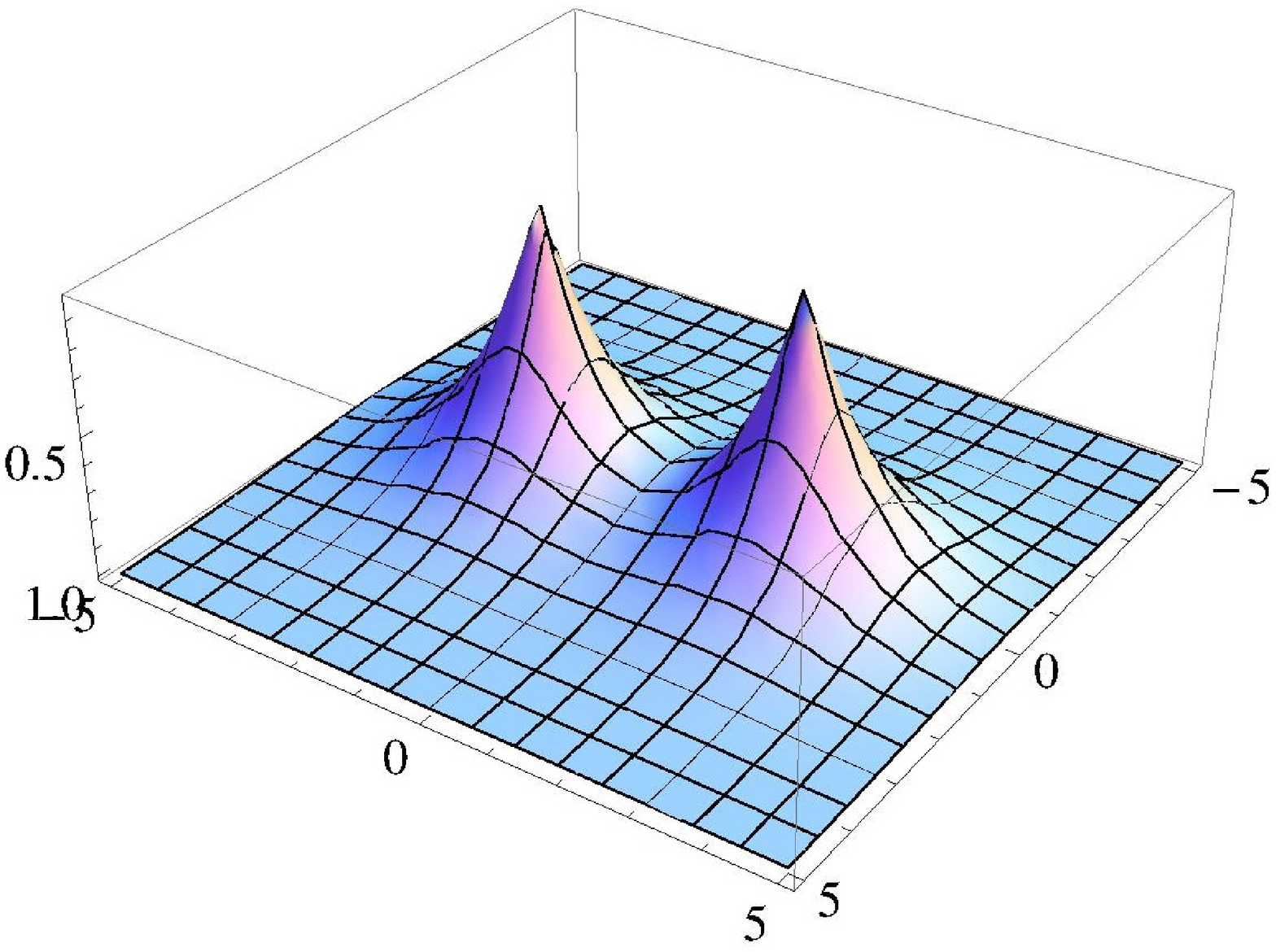}
\hspace{0.1cm}
\includegraphics[width=5cm]{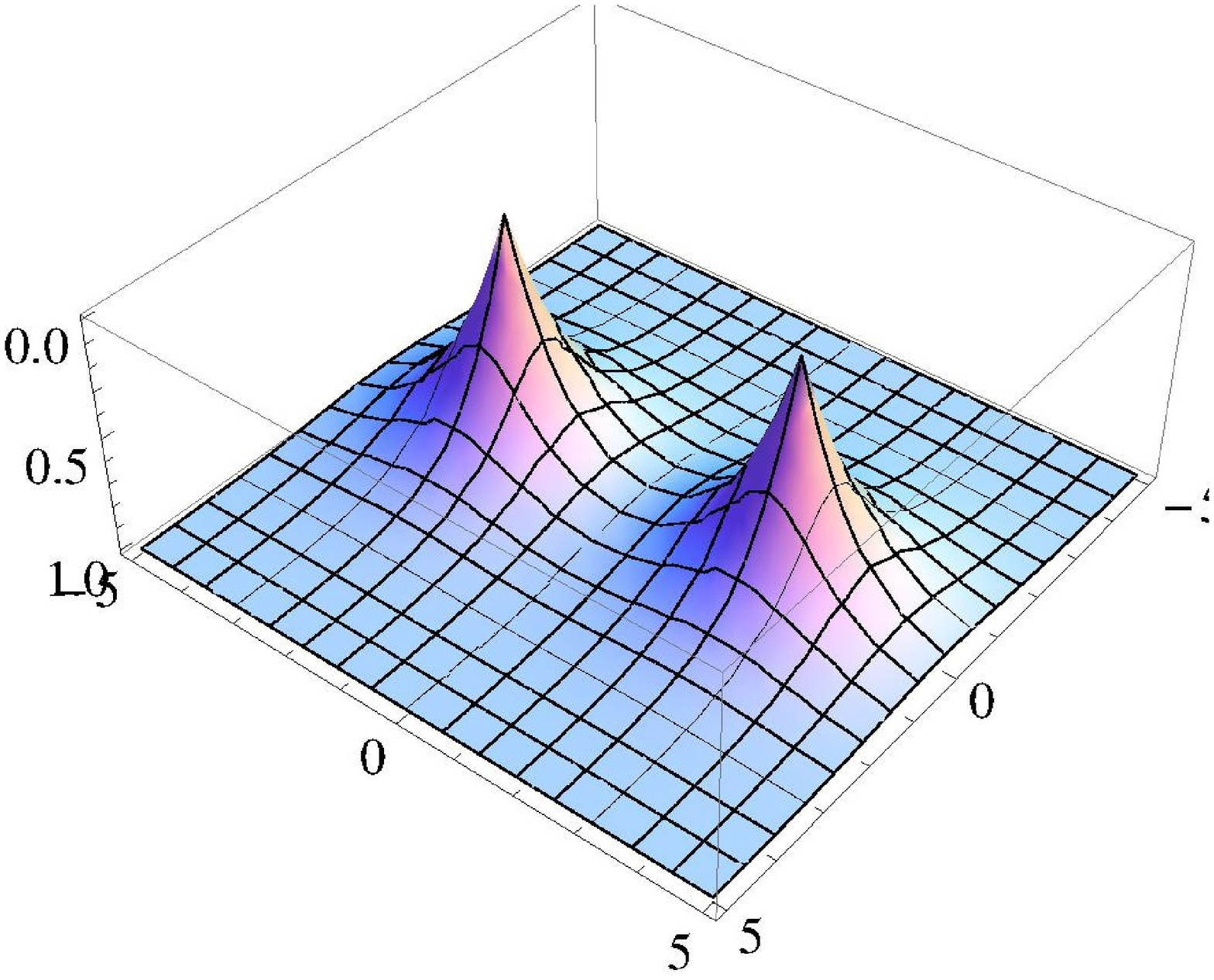}
\caption{Splitting of a $n=2$ vortex into two $n=1$ vortices.}
\label{fig4}
\end{figure}
\section{Other issues concerning the model}
As we have seen, the choices (\ref{choice}) and (\ref{poten}) for the inverse dielectric function and potential are appealing in that they provide an integrable model which makes it possible to work out analytically and in full generality the fields of self-dual topological vortices. Of course, integrable theories are always welcome because they contribute to a better understanding of the objects that they contain, but in the present case there are some aspects which could be a cause of concern, in particular the fact that the potential is infinite for zero scalar field and the divergence of the magnetic field at the center of the vortices. In this section we elaborate a little further on the theory with the aim to show that these singularities do not prevent the model from displaying a regular and well behaved phenomenology. Specifically, we will illustrate this by discussing two topics: the description of the model in terms of its elementary excitations and the interaction of fermions with vortices.
\subsection{Elementary particles and interactions}
The model (\ref{moddimen}) can be interpreted in the usual way as an effective Ginzburg-Landau theory of superconductivity, see from instance Section 21.6 of \cite{weinberg}. From this perspective, the vacuum $|\varphi|=1$ is a superconducting ground state filled with a scalar condensate of Cooper pairs originated from the interactions of some underlying microscopic theory. Instead, these interactions are weaker and Cooper pairs disappear in the normal symmetric state $\varphi=0$. The most important difference between the theory (\ref{choice})-(\ref{poten}) and the standard Abelian Higgs model is that in the latter the potential is finite for $\varphi=0$, while now $V(0)=+\infty$. This implies that, whereas in a bounded spatial domain of the AHM it would be possible to completely destroy electron pairing and to turn the field to the normal state $\varphi=0$ at a finite energetic cost (by applying, for instance, a strong external magnetic field), this is not possible within the model we are considering. The only occurrence of the symmetric phase is precisely at discrete points at the center of the vortices, where the magnetic field is allowed to become infinite inside a configuration of finite total energy. This behavior is reminiscent of Type II superconductivity, although in the present model we are at the self-dual limit and therefore, unlike in that case, vortices of any vorticity are stable, not only those with $n=1$ as in usual Type II materials. An intuitive interpretation for the large gap between the normal and superconducting phases is to suppose that (\ref{choice}) and (\ref{poten}) are suitable effective ingredients to describe a high temperature superconductor located in a thermal environment which is  well below the critical point. 

In order to study the perturbative excitations of the superconducting vacuum it is convenient, to restore dimensions coming back to the original variables given in (\ref{moddimen}). The theory is invariant under the $U(1)$ transformations 
\bdm
\phi(y)\rightarrow e^{i \omega(y)} \phi(y)\hspace{2cm}V_\alpha(y)\rightarrow V_\alpha(y)+\frac{1}{q} \frac{\partial \omega(y)}{\partial y^\alpha}
\edm
and we can use them to adopt a gauge in which the scalar field is real and positive everywhere: $\phi(y)=\rho(y)\in {\mathbb R}^+$. In this gauge the Lagrangian density is
\bdm
{\cal L}_{(\phi, V_\alpha)}=-\frac{1}{4 \mu(\frac{\rho}{v})} V_{\alpha\beta} V^{\alpha\beta} +\frac{\partial \rho}{\partial y^\alpha}\frac{\partial \rho}{\partial y_\alpha}+q^2 V_\alpha V^\alpha\rho^2-\frac{\lambda}{2}\mu(\frac{\rho}{v})(\rho^2-v^2)^2
\edm
and, with the subsequent shift of $\rho$ and rescaling of $V_\alpha$ given by
\bdm
\rho(y)=v+\frac{1}{\sqrt{2}}\eta(y)\hspace{2cm} V_\alpha=\sqrt{\mu(1)} B_\alpha,
\edm
it can be split into quadratic plus interaction parts, ${\cal L}_{(\phi, V_\alpha)}={\cal L}_{(\phi, V_\alpha)}^{(2)}+{\cal L}_{(\phi, V_\alpha)}^{\rm int}$, of the form 
\bdm
{\cal L}_{(\phi, V_\alpha)}^{(2)}=-\frac{1}{4}B_{\alpha\beta} B^{\alpha\beta} +\frac{1}{2}\frac{\partial \eta}{\partial y^\alpha}\frac{\partial \eta}{\partial y_\alpha}+q^2 v^2 \mu(1) B_\alpha B^\alpha-\lambda \mu(1) v^2 \eta^2
\edm
and 
\beq
{\cal L}_{(\phi, V_\alpha)}^{\rm int}=\sqrt{2} \mu(1) q^2 v \eta B_\alpha B^\alpha+\frac{1}{2}\mu(1) q^2 \eta^2 B_\alpha B^\alpha-\sum_{p=1}^\infty \beta_p \eta^p B_{\alpha\beta}B^{\alpha\beta}-\sum_{p=3}^\infty \gamma_p \eta^p\label{lint}.
\eeq
Here, the couplings are given in terms of derivatives of dielectric function and its inverse at $\rho=1$, and are as follows:
\beq
\beta_p= \frac{\mu(1)H^{(p)}(1)}{2^{\frac{p}{2}+2}\; p!\; v^p}\hspace{2cm}\gamma_p=\left(\lambda_{p-2}+\frac{1}{\sqrt{2}}\lambda_{p-3}+\frac{1}{8}\lambda_{p-4}\right)v^{4-p},\hspace{1cm}\lambda_p=\frac{\lambda\mu^{(p)}(1)}{2^\frac{p}{2}\; p!}\label{lintcoef},
\eeq
where we recall that $H=\frac{1}{\mu}$. Thus, for the theory (\ref{choice})-(\ref{poten}), in which $\mu(1)=\frac{1}{2}$, the spectrum consists in a massive vector boson with $M_B=q v$ and a Higgs scalar with mass $M_\eta=\sqrt{\lambda} v$. On the other hand, all the interactions can be computed from (\ref{choice}), for instance the cubic and fourth-order terms in ${\cal L}_{(\phi, V_\alpha)}^{\rm int}$ are 
\bdm
{\cal L}_{(\phi, V_\alpha)}^{\rm int\,(3)}=\frac{1}{\sqrt{2}}q^2 v \eta B_\alpha B^\alpha-\frac{1}{4\sqrt{2} v} \eta B_{\alpha\beta} B^{\alpha\beta}
\edm
\bdm
{\cal L}_{(\phi, V_\alpha)}^{\rm int\,(4)}=\frac{1}{4}q^2 \eta^2 B_\alpha B^\alpha-\frac{1}{48 v^2} \eta^2 B_{\alpha\beta} B^{\alpha\beta}-\frac{\lambda}{48}\eta^4.
\edm
Indeed, the inverse dielectric function $\mu(\rho)$ is perfectly regular at $\rho~=~1$, as one can see from Figure 1. Thus all derivatives entering in (\ref{lint})-(\ref{lintcoef}) exist and are finite and, in fact, a few explicit calculations show that their values keep decreasing as the derivative order increases. This means that the interactions between the massive fields are well defined and the model makes sense as an effective low energy theory for the elementary particles. All scattering amplitudes among them can be computed by sewing together a finite number of the tree level Feynman diagrams extracted from (\ref{lint}).
\subsection{Coupling to fermions}
In this subsection we shall consider several aspects of the physics of spin one-half fermions in the presence of a vortex. Rather than trying to present a full account of fermion dynamics under the influence of a vortex, our aim here is limited to convey some results providing evidence on the fact that the divergence of the magnetic field at the center of the vortex is not incompatible with a regular phenomenology. Thus, for simplicity, we will limit the treatment to the non-relativistic regime and will consider only the coupling of the fermions with the gauge field of the vortex, which is enough for our purposes. For a more thorough treatment of fermions on real superconductors see \cite{cdgm} and for the relativistic case with the vortex idealized as a Dirac delta flux line, see \cite{sougerb}.

Fermions of mass $M_F$ enter in the theory (\ref{moddimen}) through an additional term
\bdm
{\cal L}_{(\Psi,V_\alpha)}=\bar{\Psi}(i\gamma^\alpha \nabla_\alpha^{\cal Z}-M_F)\Psi
\edm
where the covariant derivative is $\nabla_\alpha^{\cal Z}=\frac{\partial}{\partial y^\alpha}-i {\cal Z} q V_\alpha$ and we allow for different electric charges for fermions and the scalar condensate (remember that ${\cal Z}>0$ would correspond to negative charge). The field $\Psi$ has mass-dimension $\frac{3}{2}$, an thus the rescaling 
\bdm
\Psi=q v^\frac{3}{2} \psi\hspace{3cm} M_F=q v m
\edm
gives
\bdm
{\cal L}_{(\Psi,V_\alpha)}=q^2 v^4 \left\{\bar{\psi}(i\gamma^\alpha D_\alpha^z-m)\psi \right\}
\edm
with dimensionless field $\psi$ and mass $m$, and with $D_\alpha^{\cal Z}=\partial_\alpha-i {\cal Z} A_\alpha$. The non-relativistic limit leads to the Schr\"{o}dinger-Pauli Hamiltonian
\beq
H=-\frac{1}{2m}\sum_{k=1}^2(\partial_k-i{\cal Z} A_k)^2-\frac{g {\cal Z} F_{12}}{4 m}\sigma_3 \label{hamil}
\eeq
where we have left the gyromagnetic radio $g$ unspecified to take into account the possibility that our fermions are not elementary and thus $g\neq 2$, although we do not consider anomalous magnetic moment contributions in the non-relativistic approach. The physical energies are $E_{(\Psi,V_\alpha)}=q v E$, where $E$ are the dimensionless eigenvalues of the Hamiltonian $H$, and we are treating the problem as a two-dimensional one: the energy spectrum in 3+1 dimensions is obtained by summing $\frac{p_3^2}{2m}$ to the eigenvalues obtained from $H$, and multiplying the eigenfunctions by $\frac{e^{i p_3 x^3}}{\sqrt{2\pi}}$, where $p_3$ is the momentum along the vortex axis.
\subsubsection{Zero modes and bound states}
Let us introduce the operators 
\bdm
{\cal D}=D_1^{\cal Z}-iD_2^{\cal Z}\hspace{3cm} {\cal D}^\dagger= -D_1^{\cal Z}-i D_2^{\cal Z}.
\edm
By comparing with (\ref{hamil}), we see that the matrix $H$ splits into two scalar Hamiltonians $H_\pm$, corresponding to states with spin $s_3=\pm\frac{1}{2}$, which are of the form
\beqr
H_+&=&\frac{1}{2m}{\cal D}{\cal D}^\dagger -(g-2)\frac{{\cal Z} F_{12}}{4m}\label{hmas}\\
H_-&=&\frac{1}{2m}{\cal D}^\dagger{\cal D}+(g-2)\frac{{\cal Z} F_{12}}{4m}\label{hmenos}.
\eeqr
For a cylindrically symmetric vortex we have
\beqrn
{\cal D}= e^{-i\theta}\left[\partial_r-\frac{i}{r}(\partial_\theta-i {\cal Z} A_\theta)\right]\\
{\cal D}^\dagger= -e^{i\theta}\left[\partial_r+\frac{i}{r}(\partial_\theta-i {\cal Z} A_\theta)\right]\\
\eeqrn
where $A_\theta(r)$ is given by (\ref{ansatz}) and (\ref{ader}), and $n>0$. With the explicit form of the gauge field known, it is easy to find the zero modes of these operators. For instance, if ${\cal Z}<0$ the operator ${\cal D}^\dagger$ does not have normalizable zero modes, while ${\cal D} v_l(r,\theta)=0$ for
\bdm
v_l(r,\theta)=N_l r^{{\cal Z}n-l} e^{{\cal Z} n K_0(r)}e^{i l \theta},
\edm
which can be normalized if the orbital angular momentum $l$ is $l=-(|{\cal Z}n|-2), \ldots,0$, if ${\cal Z}n$ is integer, or $l=-([|{\cal Z}n|]-1), \ldots,0$, where $[\cdot]$ is the integer part, if ${\cal Z}n$ is not integer. Thus, in these cases there are, respectively, $|{\cal Z}n|-1$ or $[|{\cal Z}n|]$ normalizable zero modes. The normalization constants are given by
\bdm
2\pi N_l^2 \int_0^\infty r^{2({\cal Z}n-l)+1} e^{2 {\cal Z} n K_0(r)}=1.
\edm
When ${\cal Z}>0$ the situation is analogous, but now zero modes of ${\cal D}$ are absent and ${\cal D}^\dagger$ has normalizable modes with the opposite signs of orbital angular momentum.
\begin{table}[t]
\begin{center}
\begin{tabular}{||c|c|c|c|c|c|c|c||}
\hline
\multicolumn{8}{||c||}{Energy $\varepsilon_l$ of bound states}\\
\hline
 &$l=0$&$l=-1$&$l=-2$&$l=-3$&$l=-4$&$l=-5$&$l=-6$\\
 \hline
$n=3$& -0.0681&  &  & & & & \\
\hline
$n=4$& -0.1513& & & & & & \\
\hline
$n=5$& -0.2470& -0.0468& & & & & \\
\hline
$n=6$& -0.3529& -0.1038& & & & & \\
 \hline
 $n=7$& -0.4674& -0.1695& -0.0390& & & & \\
 \hline
 $n=8$& -0.5892& -0.2429& -0.0858& & & & \\
 \hline
 $n=9$& -0.7175& -0.3231& -0.1396& -0.0347& & & \\
 \hline
 $n=10$& -0.8514& -0.4093& -0.1996& -0.0759& & & \\
 \hline
 $n=11$& -0.9904& -0.5008& -0.2653& -0.1229& -0.0319& & \\
 \hline
 $n=12$& -1.1340& -0.5971& -0.3360& -0.1753& -0.0695& & \\
 \hline
 $n=13$& -1.2818& -0.6978& -0.4113& -0.2325& -0.1121& -0.0300& \\
 \hline
 $n=14$& -1.4335& -0.8025& -0.4908& -0.2942& -0.1594& -0.0649& \\
 \hline
 $n=15$& -1.5887& -0.9110& -0.5742& -0.3599& -0.2110& -0.1043& -0.0284\\
 \hline
\end{tabular}
\end{center}
\caption{Energy $\varepsilon_l$ of bound states for a proton in the vortex field for several values of the orbital angular momentum $l$ and vorticity $n$.}
\label{table1}
\end{table}

In view of (\ref{hmas})-(\ref{hmenos}), if the fermion is an elementary particle the zero modes of ${\cal D}^\dagger$ and ${\cal D}$ are zero-energy solutions of the Schr\"{o}dinger-Pauli equation for, respectively, spin up or down. Instead, they can become bound states for the case of composite fermionic particles with gyromagnetic ratio different from two. For instance, if we deal with a fermion of positive electric charge and $g>2$, such as a proton, the modes $v_l(r,\theta)$ adquire negative energy which, if the constants of the theory are such that the coefficient of the second term in (\ref{hmas}) is small, can be computed in first order perturbation theory:
\beq
E_l=\frac{(g-2) {\cal Z}}{4m}\int d^2x F_{12}|v_l|^2=\frac{\pi(g-2) {\cal Z} n}{2m} N_l^2\int_0^\infty dr r^{2({\cal Z}n-l)+1} K_0(r) e^{2 {\cal Z} n K_0(r)}.\label{boundfer}
\eeq
For concreteness, we can take $q=2e$ for a condensate of Cooper pairs, and then ${\cal Z}=-\frac{1}{2}$ for the proton. Thus, according with the previous analysis of zero modes, we conclude that:
\begin{itemize}
\item If the vorticity $n$ is even, $H_-$ has $\frac{n}{2}-1$ bound states corresponding to orbital angular momentum $l=0,-1,\ldots,-\left(\frac{n}{2}-2\right)$.
\item If the vorticity $n$ is odd , $H_-$ has $\frac{n-1}{2}$ bound states corresponding to orbital angular momentum $l=0,-1,\ldots,-\frac{n-3}{2}$.
\end{itemize} 
The energy of some of these bound states, in the form $E_l=\frac{\pi(g-2)}{2m}\varepsilon_l$ and computed numerically using (\ref{boundfer}) are given in Table 1. We see that they are finite and are not affected by the divergence of the magnetic field at the origin, which is overcome by the rate at which the wave function vanishes at this point.
\subsubsection{Scattering states and phase shifts}
Let us now turn to scattering states. Changing to polar gauge field components and using (\ref{ansatz}) in (\ref{hamil}), the Schr\"{o}dinger-Pauli equation is
\bdm
\left[\partial_r^2+\frac{1}{r}\partial_r+\frac{1}{r^2}(\partial_\theta-i {\cal Z} A_\theta)^2+\frac{g{\cal Z}}{2r}\frac{dA_\theta}{dr}\sigma_3+k^2\right]\psi=0,\hspace{1cm}k=\sqrt{2mE},
\edm
where now the energy $E$ is positive. In particular, for states of orbital angular momentum $l$
\bdm
\psi(r,\theta)=\xi_l(r) e^{i l\theta}
\edm
the equation takes the form
\bdm
r^2 \frac{d^2\xi_l}{dr^2}+r \frac{d\xi_l}{dr}+\left[ k^2 r^2-(l-{\cal Z} A_\theta)^2+\frac{1}{2} g {\cal Z} r \frac{dA_\theta}{dr}\sigma_3\right]\xi_l=0.
\edm
Since the last term is proportional to the the product $r^2 F_{12}(r)$, the divergence of the magnetic field at the origin is harmless. On the other hand, very far from the origin, where $A_\theta\simeq n$ and the magnetic field vanishes at a exponential rate, the solution for each spin component is a linear combination of Bessel functions
\beq
\psi_l(r,\theta)=\left(a_l J_{|l-{\cal Z}n|}(kr)+b_l Y_{|l-{\cal Z}n|}(kr)\right)e^{i l \theta}\hspace{1.5cm} {\rm for}\ r\gg 1,\label{solasymp}
\eeq
where both coefficients $a_l$ and $b_l$ can be taken to be real. Except for an arbitrary global normalization, they are fixed by imposing regularity at the origin once the solution is extended to the whole plane. Using the expressions of the Bessel functions valid for great $r$
\beqr
J_m(z)&\simeq& \sqrt{\frac{2}{\pi z}} \cos\left(z-m\frac{\pi}{2}-\frac{\pi}{4}\right)\label{bessasymp1}\\
Y_m(z)&\simeq& \sqrt{\frac{2}{\pi z}} \sin\left(z-m\frac{\pi}{2}-\frac{\pi}{4}\right)\label{bessasymp2}
\eeqr
the asymptotic wave function (\ref{solasymp}) can be written as
\beq
\psi_l(r,\theta)=\frac{1}{\sqrt{2\pi k}}\left(\frac{e^{i k r}}{\sqrt{r}} e^{-i\frac{\pi}{2}|l-{\cal Z}n|}e^{-i\frac{\pi}{4}} (a_l-i b_l)+\frac{e^{-i k r}}{\sqrt{r}} e^{i\frac{\pi}{2}|l-{\cal Z}n|}e^{i\frac{\pi}{4}} (a_l+i b_l)\right) e^{i l \theta}.\label{asympvort}
\eeq
For a free fermion, without magnetic field or potential, the asymptotic linear combination of Bessel functions (\ref{solasymp}) solves the Schr\"{o}dinger-Pauli equation for all $r$, but $Y_{|l|}(kr)$ blows up at the origin. Thus the free solution with orbital angular momentum $l$ is
\bdm
\psi_l^{\rm free}(r,\theta)=J_{|l|}(kr)e^{i l \theta}
\edm
and, using (\ref{bessasymp1}) again, we can write this free wave function as
\beq
\psi_l^{\rm free}(r,\theta)=\frac{1}{\sqrt{2\pi k}}\left(\frac{e^{i k r}}{\sqrt{r}} e^{-i\frac{\pi}{2}|l|}e^{-i\frac{\pi}{4}}+\frac{e^{-i k r}}{\sqrt{r}} e^{i\frac{\pi}{2}|l|}e^{i\frac{\pi}{4}} \right) e^{i l \theta}.\label{freesol}
\eeq
We can imagine (\ref{freesol}) as a superposition of two circular waves, one incoming from great distance to the origin and the other being scattered from it towards infinity with a phase change. When an interaction with cylindrical symmetry is at work, due to the join conservation of angular momentum and probability, we expect for the asymptotic solution the same structure of (\ref{freesol}), with the incoming and  outgoing waves having the same amplitude but a different phase, i.e., we should have something like
\beq
\psi_l(r,\theta)=A\left(\frac{e^{i( k r+2\delta_l)}}{\sqrt{r}} e^{-i\frac{\pi}{2}|l|}e^{-i\frac{\pi}{4}}+\frac{e^{-i k r}}{\sqrt{r}} e^{i\frac{\pi}{2}|l|}e^{i\frac{\pi}{4}}\right) e^{i l \theta}.\hspace{1.5cm} {\rm for}\ r\gg 1,\label{deltal}
\eeq
where $\delta_l$ is the phase shift due to the interaction. Thus,  by comparing (\ref{asympvort}) with (\ref{deltal}), we find for the phase shifts produced by the vortex magnetic field the following expression:
\beq
\delta_l=\frac{\pi}{2}(|l|-|l-{\cal Z}n|)-\arctan(\frac{b_l}{a_l}).\label{phaseshift}
\eeq
The first term is precisely the Aharonov-Bohm result $\delta_l^{AB}=\frac{\pi}{2}(|l|-|l-{\cal Z}n|)$ \cite{ahboh, hagen}, corresponding to a situation in which the magnetic field is confined within an infinitesimally thin tube along the $x^3$ axis. For ${\cal Z}n\in{\mathbb Z}$ this reduces to $e^{2 i \delta_l^{AB}}=(-1)^{{\cal Z}n}$, i.e., the contribution to the phase shift factor in (\ref{deltal}) is +1 or -1, but the same for all $l$. This means that if we take a superposition of circular incoming waves with different $l$ values, we obtain the same superposition of outgoing circular waves than in free case, except for a global physically irrelevant factor. Thus, as it is well known, an infinitesimally thin flux tube with integer $\frac{\Phi_M}{2\pi}$ has not physical effects. The situation is different for the vortex: in this case the integer flux of $\frac{\Phi_M}{2\pi}$ is not innocuous due to the second term in (\ref{phaseshift}). As an illustration, we have solved numerically the Schr\"{o}dinger-Pauli equation to compute some of the phase shifts produced by the vortex with $n=1$. In Table 2 we give the phase-shifts for the spin up states of an electron (${\cal Z}=\frac{1}{2}, g=2$) for several values of the momentum $k$ and angular momentum $l$, and in Figure 5 we show how the phase-shifts change with $l$ for a fixed energy. As one can see, the results are perfectly regular, with phase-shifts decreasing with energy and showing also a dependence with $l$ which reflects the non-trivial interaction produced by the vortex as compared to the Aharonov-Bohm background.
\begin{table}
\begin{center}
\begin{tabular}{||c|c|c|c|c|c||}
\hline
\multicolumn{6}{||c||}{Phase shifts for $n=1$}\\
\hline
 &$l=-2$&$l=-1$&$l=0$&$l=1$&$l=2$\\
 \hline
 $k=1$& -0.6711& -0.4438& 0.2619& 0.5595& 0.6918\\
 \hline
 $k=2$& -0.4666& -0.2367& 0.1455& 0.3794& 0.5288\\
 \hline
 $k=3$& -0.3452& -0.1587& 0.0999& 0.2843& 0.4191\\
 \hline
 $k=4$& -0.2722& -0.1191& 0.0758& 0.2267& 0.3452\\
 \hline
 $k=5$& -0.2242& -0.0953& 0.0610& 0.1883& 0.2929\\
 \hline
 $k=6$& -0.1905& -0.0794& 0.0510& 0.1610& 0.2541\\
 \hline
 $k=7$& -0.1656& -0.0680& 0.0438& 0.1405& 0.2243\\
 \hline
 $k=8$& -0.1464& -0.0595& 0.0384& 0.1247& 0.2006\\
 \hline
 $k=9$& -0.1312& -0.0529& 0.0342& 0.1120& 0.1815\\
 \hline
 $k=10$& -0.1189& -0.0475& 0.0308& 0.1017& 0.1656\\
 \hline
\end{tabular}
\end{center}
\caption{Phase shifts for the scattering of an electron with spin up from the $n=1$ vortex.}
\label{table2}
\end{table}
\begin{figure}
\centering
\includegraphics[width=7cm]{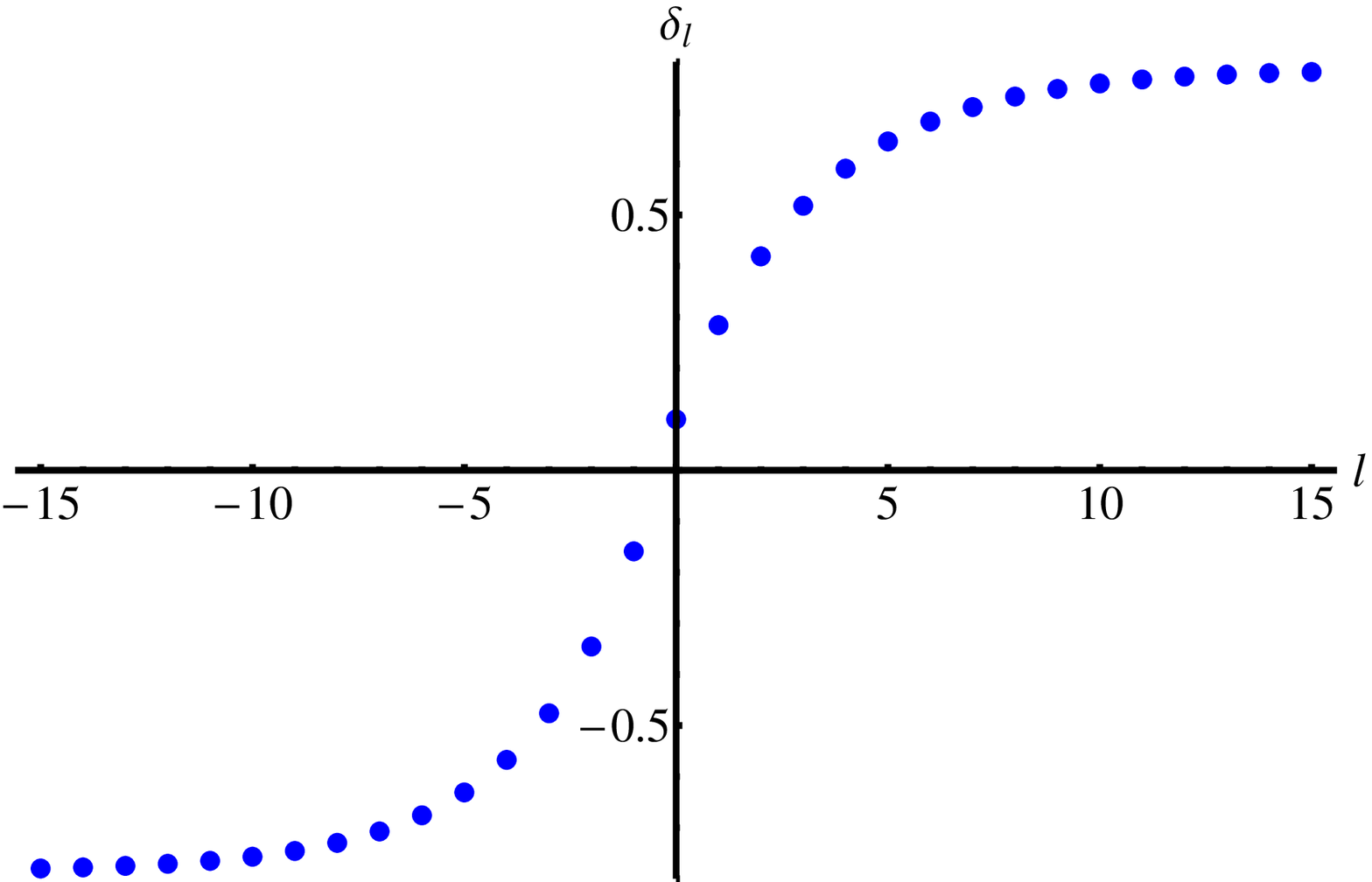}
\hspace{0.2cm}
\includegraphics[width=7cm]{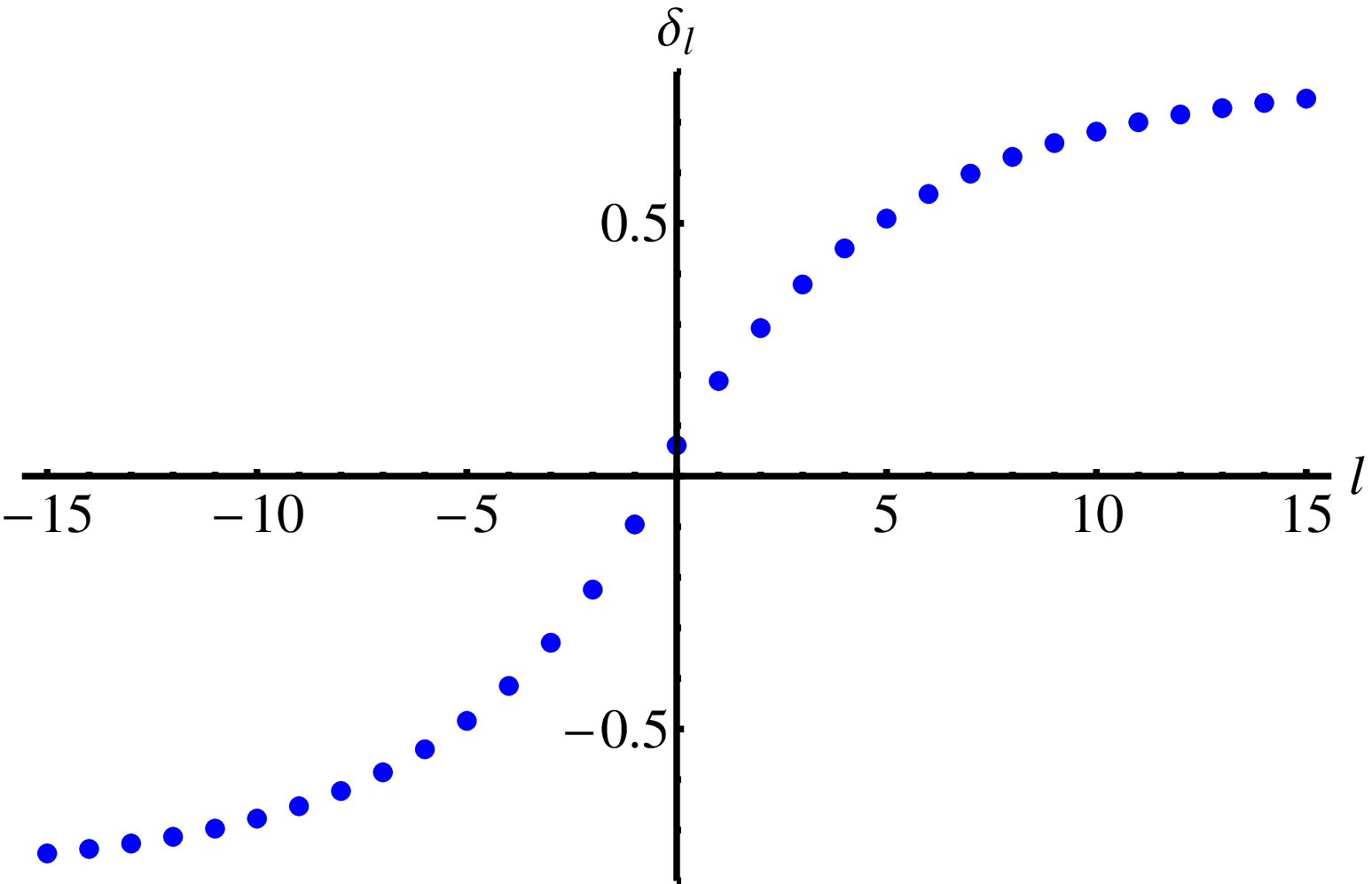}
\caption{Phase shifts of an electron with spin up scattered from the $n=1$ vortex as a function of orbital angular momentum for $k=3$ (left) and $k=5$.}
\label{fig5}
\end{figure}
\section{Conclusions}
In this note, we have examined a generalized Abelian Higgs model in which the solutions for topological vortices of any magnetic flux $\Phi_M=2\pi n, n\in{\mathbb Z}$, are analytical. The model comes about by reducing Bogomolny equations for a cylindrically symmetric vortices to a second order ODE, and then looking for a form of the dielectric function which makes the ODE linear. The solution of the ODE, given in terms of Bessel functions, can be made compatible with the boundary conditions required for finite energy. The dielectric function and potential are regular, except when the scalar field goes to zero. Despite the singularity appearing at this point, the model can be understood as a well behaved effective theory describing the low energy interactions of a massive vector boson and a Higgs field which arise perturbatively around a vacuum which engenders symmetry breaking. Once the solutions for cylindrically symmetric vortices are found, their structure gives a strong clue that the solutions corresponding to separated multivortices must be given by exact analytical expressions too, and a more detailed analysis confirms this. A salient feature of the solutions is that, at the center of the vortices, both the magnetic field and the energy density diverge. Nevertheless, the divergences are mild enough to make the total magnetic flux and energy finite and consistent with the Bogomolny bound. The divergences do not spoil either other facets of the theory, like the dynamics of fermions on the vortex field, which display a regular phenomenology.
\section*{Acknowledgments}
This study was partially supported by MCIN with funding from European Union NextGenerationEU (PRTR-C17.I1) and Consejer\'{\i}a de Educaci\'on from JCyL through QCAYLE project, as well as PID2020-113406GB-I0 project by MCIN of Spain.

\end{document}